\documentclass[aps,prd,reprint]{revtex4-2}

\usepackage[english]{babel}
\usepackage{amsfonts}
\usepackage{amsmath}
\usepackage{graphicx}
\usepackage{mathtools}
\usepackage{multirow}
\newcommand{\C}[1]{$^{\circ}$C}
\usepackage{color}
\usepackage{xcolor}

\begin{document}

\title{Comprehensive study of amorphous metal oxide and Ta$_2$O$_5$-based mixed oxide coatings for gravitational-wave detectors}

\author{Mariana A. Fazio}
\email{Mariana.Fazio@colostate.edu}
\affiliation{Department of Electrical and Computer Engineering, Colorado State University, Fort Collins, CO, USA}

\author{Gabriele Vajente}
\affiliation{LIGO Laboratory, California Institute of Technology, Pasadena, CA, USA}

\author{Le Yang}
\affiliation{Department of Chemistry, Colorado State University, Fort Collins, CO, USA}

\author{Alena Ananyeva}
\affiliation{ LIGO Laboratory, California Institute of Technology, Pasadena, CA, USA}

\author{Carmen S. Menoni}
\email{Carmen.Menoni@colostate.edu}
\affiliation{Department of Electrical and Computer Engineering, Colorado State University, Fort Collins, CO, USA}
\affiliation{Department of Chemistry, Colorado State University, Fort Collins, CO, USA}
\date{\today}

\begin{abstract}

High finesse optical cavities of current interferometric gravitational-wave detectors are significantly limited in sensitivity by laser quantum noise and coating thermal noise. The thermal noise is associated with internal energy dissipation in the materials that compose the test masses of the interferometer. Our understanding of how the internal friction is linked to the amorphous material structure is limited due to the complexity of the problem and the lack of studies that span over a large range of materials. We present a systematic investigation of amorphous metal oxide and Ta$_2$O$_5$-based mixed oxide coatings to evaluate their suitability for low Brownian noise experiments. It is shown that the mechanical loss of metal oxides is correlated to their amorphous morphology, with continuous random network materials such as SiO$_2$ and GeO$_2$ featuring the lowest loss angles. We evaluated different Ta$_2$O$_5$-based mixed oxide thin films and studied the influence of the dopant in the optical and elastic properties of the coating. We estimated the thermal noise associated with high-reflectance multilayer stacks that employ each of the mixed oxides as the high index material. We concluded that the current high index material of TiO$_2$-doped Ta$_2$O$_5$ is the optimal choice for reduced thermal noise among Ta$_2$O$_5$-based mixed oxide coatings with low dopant concentrations.

\end{abstract}

\maketitle

\section{Introduction}

Gravitational-wave (GW) detectors such as Advanced LIGO \cite{abbott2016gw150914}, Advanced Virgo \cite{acernese2014advanced} and KAGRA \cite{akutsu2015large} are interferometric devices that utilize high finesse optical cavities to perform high precision displacement measurements. These impressive km-long facilities allow the study of astrophysical phenomena but their sensitivity is limited by much smaller scale physics, mainly laser quantum noise \cite{buonanno2001quantum} and Brownian motion of the end test masses. This Brownian motion arises from internal energy dissipation in the materials composing the test masses of the interferometer. This dissipation arises from Brownian motion noise that leads to thermally driven optical path length variations \cite{saulson1990thermal, numata2004thermal} and also affects other high precision experiments such as optical clocks \cite{ludlow2008sr, rosenband2008frequency, bishof2013optical}.

In GW detectors, the test masses mirrors consists of a high purity SiO$_2$ substrate coated with a high reflectivity multilayer stack of amorphous oxide thin films. This multilayer stack is composed of alternating layers of low and high refractive index materials, with LIGO and Virgo originally employing SiO$_2$ as the low index material and Ta$_2$O$_5$ as the high index material. Later, Advanced LIGO and Advanced Virgo used a  mixture of TiO$_2$ and Ta$_2$O$_5$ as the high index material \cite{granata2020amorphous}. For these interferometric experiments, the power spectral density of Brownian noise (or thermal noise) at a frequency $f$ can be expressed as \cite{martin2012coating}:

\begin{equation}
S_{Brownian}(f)=\frac{2k_BT}{\pi^2f}\frac{d \phi}{w^2Y_S} \left(\frac{Y_C}{Y_S}+\frac{Y_S}{Y_C}\right)
\label{eq:thermalnoise}
\end{equation}

where $k_B$ is the Boltzmann's constant, $T$ is the temperature, $w$ is the diameter of the laser beam probing the multilayer stack motion, $Y_S$ and $Y_C$ are the Young's modulus of the substrate and the multilayer stack respectively, $\phi$ is the mechanical loss angle of the stack and $d$ the total thickness of the stack. This expression assumes that bulk and shear loss angles are equal and that the Poisson ratios for all materials are negligible. It also requires an approximation to define the Young modulus and loss angle of the multilayer stack, for which an effective medium approach can be employed. Therefore, the Young modulus and loss angle of the multilayer stack are defined as the average of the values corresponding to each composing material (for the current GW detectors, SiO$_2$ and TiO$_2$:Ta$_2$O$_5$ mixed oxide) weighted by the elastic energy stored in them \cite{fejer2021effective}. The expression \ref{eq:thermalnoise} requires several approximations, however it is a valuable tool to evaluate overall scaling behavior and compare between different compositions for the multilayer stacks. It also clearly indicates that the different material properties of the layers, such as refractive index, Young's modulus and loss angle, play a major role in the resulting thermal noise of the multilayer stack. Therefore optical materials should be fully characterized to determine their suitability for use in GW detectors or other sensitive devices in which the thermal noise can be a limiting factor.

The major source of thermal noise in the multilayer stacks originally employed by LIGO and Virgo was found to be in the Ta$_2$O$_5$ layers with only a minor contribution associated to the SiO$_2$ layers \cite{penn2003mechanical}. Advanced LIGO and Advanced Virgo replaced the high index layer in the stacks with TiO$_2$:Ta$_2$O$_5$ with a cation ratio of 0.27 after finding that the addition of TiO$_2$ reduced the mechanical losses by around 40\% \cite{abbott2016gw150914,granata2016mechanical,granata2020amorphous}. The cation ratio is defined as M/(M +Ta) with M the dopant atomic concentration (in this case titanium) and Ta the tantalum atomic concentration. This stimulated research efforts directed at identifying the reason of this reported decrease caused by TiO$_2$ doping of Ta$_2$O$_5$ and also at exploring other oxide mixtures of high index that might featured low mechanical loss.

The effect of the addition of TiO$_2$ is complex, inducing changes not only in the optical properties of the material but also structural and morphological modifications. It was found that the cation ratio is a key parameter that has a profound impact on the mechanical loss of TiO$_2$:Ta$_2$O$_5$ \cite{harry2006titania}. Transmission electron microscopy studies of low mechanical loss TiO$_2$:Ta$_2$O$_5$ mixed films with cation ratios around 0.2 - 0.3 showed that the dopant promotes structural homogeneity at the nearest-neighbor level and might also prevent oxygen loss \cite{bassiri2013correlations}. Characterization of the structure, morphology and optical properties of TiO$_2$:Ta$_2$O$_5$ mixed films with a cation ratio of 0.27 that featured low mechanical loss found that the dopant increased the crystallization temperature, lowered the optical absorption loss and induced the formation of a ternary compound in the crystallized film \cite{fazio2020structure}. It remains unclear how these specific features of the material contribute to lowering the mechanical losses, but they can prove to be useful in guiding the design of new materials targeted to low Brownian noise applications. Several studies have also focused on Ta$_2$O$_5$ mixed with other oxides, such as CoO:Ta$_2$O$_5$ and WO$_2$:Ta$_2$O$_5$ \cite{flaminio2010study}, Sc$_2$O$_3$:Ta$_2$O$_5$ \cite{fazio2020growth}, SiO$_2$:Ta$_2$O$_5$ \cite{yang2020structural} and ZrO$_2$:Ta$_2$O$_5$ \cite{abernathy2021exploration}. Extensive research has also been conducted on the mechanical loss of other oxides and mixtures such as TiO$_2$:Nb$_2$O$_5$ \cite{granata2020progress, PhysRevD.103.072001} and a few ZrO$_2$-based mixtures \cite{flaminio2010study}. However no systematic studies have been carried out that compare suitable Ta$_2$O$_5$-based mixtures to discern the effect the dopant has on mechanical loss. In addition, there is a notable lack of information in the literature on the elastic properties and mechanical loss of other well-known oxides. Reporting on materials and properties that correlate not only to low losses but also to higher losses is critical for building a foundation of knowledge on the origins of mechanical loss in oxide thin films that can fuel future research on novel materials for GW detectors.

Herein we conduct a comprehensive study of different amorphous metal oxide and Ta$_2$O$_5$-based mixed oxide coatings for use in current GW detectors. It is shown that the mechanical loss of metal oxides is correlated to their amorphous morphology, with continuous random network materials such as SiO$_2$ and GeO$_2$ featuring the lowest loss angles. We evaluated different Ta$_2$O$_5$-based mixed oxide thin films and studied the influence of the dopant in the optical and elastic properties of the coating. We estimated the thermal noise associated with high-reflectance multilayer stacks that employ each of the mixed oxides as the high index material. We concluded that among Ta$_2$O$_5$-based mixed oxide coatings with low dopant concentrations the current high index material of TiO$_2$-doped Ta$_2$O$_5$ is the optimal choice for low thermal noise.

\section{Experimental} \label{sec:experimental}

Thin films were grown by reactive ion beam sputtering employing the Laboratory Alloy and Nanolayer System (LANS) manufactured by 4Wave, Inc \cite{zhurin2000biased}. Details of the deposition technique can be found in \cite{fazio2020growth}. In this study, metallic targets of Ta, Al, Si, Sc, Ti, Zn, Zr, Nb, Y, Ge and Hf of 99.99\% purity were employed. The target pulse period was fixed to be 100 $\mu$s and the oxygen flow and pulse width were varied to obtain near stoichiometric films. Deposition conditions for oxide films are presented in Table \ref{table:dep-oxides} along with their corresponding deposition rates. For Ta$_2$O$_5$-based mixed oxide coatings, the pulse width was adjusted to obtain a dopant cation ratio around 0.2. The oxygen flow was set at 12 sccm which was sufficient to achieve the desired stoichiometry. The deposition conditions for the mixed oxide films are presented in Table \ref{table:dep-doped} with their corresponding deposition rates and dopant cation ratio determined from x-ray photoelectron spectroscopy (XPS) atomic concentrations. The films were grown on 25.4 mm diameter and 6.35 mm thick ultraviolet grade fused silica substrates and on Si (100) wafers. For mechanical loss measurements coatings were deposited on 75 mm diameter and 1 mm thick fused silica substrates. Post deposition annealing in air was carried out using a heating rate of 100\C{} per hour and samples were soaked for 10 hours at 300\C{}, 500\C{}, 600\C{}, 700\C{}, 800\C{} and 900\C{} until crystallization was reached.

\begin{table*}[!ht]
	\setlength\tabcolsep{5pt}
	\begin{center}
		\begin{tabular}{ccccc}
			\hline
			Material & Target & Pulse width ($\mu$s) & O$_2$ flow (sccm) & Deposition rate (nm/s)\\
			\hline
			Ta$_2$O$_5$ & Ta & 2 & 14 & 0.0209 $\pm$ 0.0001\\
			Nb$_2$O$_5$ & Nb & 2 & 14 & 0.0149 $\pm$ 0.0001\\
			Al$_2$O$_3$ & Al & 2 & 10 & 0.00959 $\pm$ 0.00005\\
			Y$_2$O$_3$ & Y & 2 & 14 & 0.0225 $\pm$ 0.0001\\
			ZrO$_2$ & Zr & 2 & 12 & 0.0116 $\pm$ 0.0001 \\
			HfO$_2$ & Hf & 2 & 14 & 0.0135 $\pm$ 0.0001 \\
			SiO$_2$ & Si & 50 & 3 & 0.0152 $\pm$ 0.0001\\
			GeO$_2$ & Ge & 50 & 6 & 0.0110 $\pm$ 0.0001\\
			\hline
		\end{tabular}
	\end{center}
	\caption{Deposition conditions for the metal oxide coatings evaluated in this study. In all cases, the oxygen flow was set at 12 sccm.}
	\label{table:dep-oxides}
\end{table*}

\begin{table*}[!ht]
	\setlength\tabcolsep{5pt}
	\begin{center}
		\begin{tabular}{ccccc}
			\hline
			Material & Targets & Pulse width ($\mu$s) & Deposition rate (nm/s) & Dopant cation ratio\\
			\hline
			Al$_2$O$_3$:Ta$_2$O$_5$ & Al - Ta & 51 - 2 & 0.0271 $\pm$ 0.0001 & 0.17 $\pm$ 0.01\\
			SiO$_2$:Ta$_2$O$_5$ &  Si - Ta & 72 - 2 & 0.0246 $\pm$ 0.0001 & 0.26 $\pm$ 0.01\\
			Sc$_2$O$_3$:Ta$_2$O$_5$ & Sc - Ta & 45 - 2 & 0.0287 $\pm$ 0.0001 & 0.105 $\pm$ 0.007\\
			TiO$_2$:Ta$_2$O$_5$ & Ti - Ta & 2 - 53 & 0.01603 $\pm$ 0.00005 & 0.27 $\pm$ 0.04\\
			ZnO:Ta$_2$O$_5$ & Zn - Ta & 56 - 2 & 0.0254 $\pm$ 0.0001 & 0.20 $\pm$ 0.01\\
			Y$_2$O$_3$:Ta$_2$O$_5$ &  Y - Ta & 73 - 2 & 0.0282 $\pm$ 0.0001 & 0.24 $\pm$ 0.02\\
			ZrO$_2$:Ta$_2$O$_5$ & Zr - Ta & 54 - 2 & 0.0294 $\pm$ 0.0001 & 0.23 $\pm$ 0.01\\
			Nb$_2$O$_5$:Ta$_2$O$_5$ & Nb - Ta & 68 - 2 & 0.0261 $\pm$ 0.0001 & 0.12 $\pm$ 0.01\\
			HfO$_2$:Ta$_2$O$_5$ & Hf - Ta & 49 - 2 & 0.0295 $\pm$ 0.0001 & 0.23 $\pm$ 0.02\\
			\hline
		\end{tabular}
	\end{center}
	\caption{Deposition conditions for the Ta$_2$O$_5$-based mixed oxide coatings in this study along with cation ratios determined from XPS atomic concentrations.}
	\label{table:dep-doped}
\end{table*}

 
The optical characterization of the films involved spectroscopic ellipsometry and measurements of absorption loss at $\lambda$ = 1064 nm. Ellipsometric data were collected using a Horiba UVISEL ellipsometer in a spectral range of 0.59 eV to 6.5 eV at an angle of incidence of 60$^{\circ}$. Fitting of spectroscopic ellipsometry data with the DeltaPsi2 software provided estimates of the film thickness and optical parameters, in particular the refractive index at the wavelength of interest for the application of the coatings ($\lambda$ = 1064 nm). Absorption loss was measured by photothermal common path interferometry \cite{alexandrovski2009photothermal} at $\lambda$ = 1064 nm. Five spots on the surface of each sample were measured in a 4 mm $\times$ 4 mm area.

XPS measurements to determine atomic concentration, and hence dopant cation ratio, on the mixed oxide films were carried out with a Physical Electronics PE 5800 ESCA/ASE system equipped with a 2 mm monochromatic Al K$\alpha$ x-ray source, a hemispherical electron analyzer and a multichannel detector. The photoelectron take-off angle was 45$^{\circ}$ and we employed a charge neutralizer with a current of 10 $\mu A$ for all measurements. The instrument base pressure was around $1\times10^{-9}$ Torr. Survey spectra were collected at a pass energy of 188 eV, 0.8 eV step, 30 s sweep intervals and multiple sweeps were taken to achieve a reasonable peak-to-noise ratio. XPS spectra were analyzed with the CasaXPS software (version 2.3.19) \cite{casaxps}. Grazing incidence x-ray diffraction (GIXRD) was employed to determine whether the films remained amorphous or crystallized after each annealing step. These measurements were performed using a Bruker D8 Discover Series I diffractometer with a Cu K$\alpha$ source. The incident angle was fixed at $0.5^{\circ}$ and $2\theta$ was varied between $10^{\circ}$ and $80^{\circ}$.

The mechanical properties of the film were measured using the 75-mm-diameter and 1-mm-thickness disks. The samples were suspended in a Gentle Nodal Suspension \cite{cesarini2009, vajente2017}: the disk is sitting on the top of a semi-spherical support and is kept in place only due to gravity and static friction. The system is housed inside a vacuum chamber with residual gas pressure below $10^{-6}$ Torr. In this way the disk acts as a high-quality-factor mechanical resonator, where about 20 modes with frequencies ranging from 1 kHz to 30 kHz can be probed, virtually free of recoil energy loss due to the suspension point contact.

All substrates were measured before the film was deposited, and after an initial heat treatment at 900$^\circ$C for 10 hours. The quality factor $Q_i^{\mathrm{(sub)}}$ and frequency $f_i^{\mathrm{(sub)}}$ of all modes were measured, to provide a reference for the measurements of the coated disks. The measurement is performed by exciting all the disk resonant modes simultaneously with an electrostatic drive \cite{vajente2017} and then tracking the mode amplitude decay over time. The quality factor, defined as the number of oscillations after which the amplitude of a given mode decays by $1/e$, can be determined from the time evolution of the observed mode amplitudes. After deposition of the films, the samples were measured again, obtaining a new set of quality factors $Q_i^{\mathrm{(coat})}$ and mode frequencies $f_i^{\mathrm{(coat})}$. The presence of the thin film has two effects: the film changes the flexural rigidity \cite{rao2007} of the disk and therefore induces a shift of all resonant frequencies; the film introduces additional energy loss mechanisms due to the material loss angle, reducing the quality factor of each mode.

The change in the resonant mode frequencies $\Delta f_i = f_i^{\mathrm{(coat})} - f_i^{\mathrm{(sub})}$ is a function of the elastic properties of the film: Young's modulus, Poisson ratio, density and thickness. The thickness of the film was obtained by ellipsometry, and the density was estimated from literature values. The mode frequency shifts measured experimentally are then fitted to a finite element model \cite{granata2020amorphous, vajente2020}, using a Bayesian approach, to obtain an estimate of the Young's modulus and Poisson ratio. Since the energy loss mechanisms in the substrates and films are independent (we are assuming there are no interface effects), the loss angle of the film is related to the change in the measured quality factor by a simple relation
\begin{equation}
    \phi_i^{\mathrm{(coat)}} = \frac{E_i^{\mathrm{(sub)}}}{E_i^{\mathrm{(film)}}+E_i^{\mathrm{(sub)}}} \phi_i^{\mathrm{(sub)}} + \frac{E_i^{\mathrm{(film)}}}{E_i^{\mathrm{(film)}}+E_i^{\mathrm{(sub)}}} \phi_i^{\mathrm{(film)}}
\end{equation}
where we have defined the loss angle as the inverse of the quality factor $\phi = Q^{-1}$ and denoted with $E_i$ the elastic energy stored on average in the $i$-th resonant mode, for either the substrate or the film. Those energies can be computed with a finite element analysis, using the estimated values of Young's modulus and Poisson ratio, and then used in the above equation to extract the film material loss angle. The loss angles showed minimal variations with frequency, thus average values at 1 kHz are reported. 

In the case of the mixed oxide coatings, the thermal noise was estimated for an HR stack composed of layers of silica and the doped tantala material. Using the refractive index of each mixed material, a multilayer HR stack was designed with a target transmission of 5 ppm at $\lambda$ = 1064 nm which corresponds to the end test mass transmission requirement. The design is obtained by starting with a stack of doublets where each layer has an optical thickness of one quarter wavelength. The number of doublets is chosen to get a transmission as close as possible to 5 ppm. Once the design is obtained, the coating Brownian thermal noise is computed using the measured elastic properties and the model developed by Hong \textit{et. al.} \cite{Hong2013}. This allows for direct comparison of the performance of the mixed oxide coatings, taking into account not only their loss angles but also their elastic and optical properties.

\section{Results} \label{sec:results}

\subsection{Oxide coatings}

\begin{figure}[h!]
	\includegraphics[width=\linewidth]{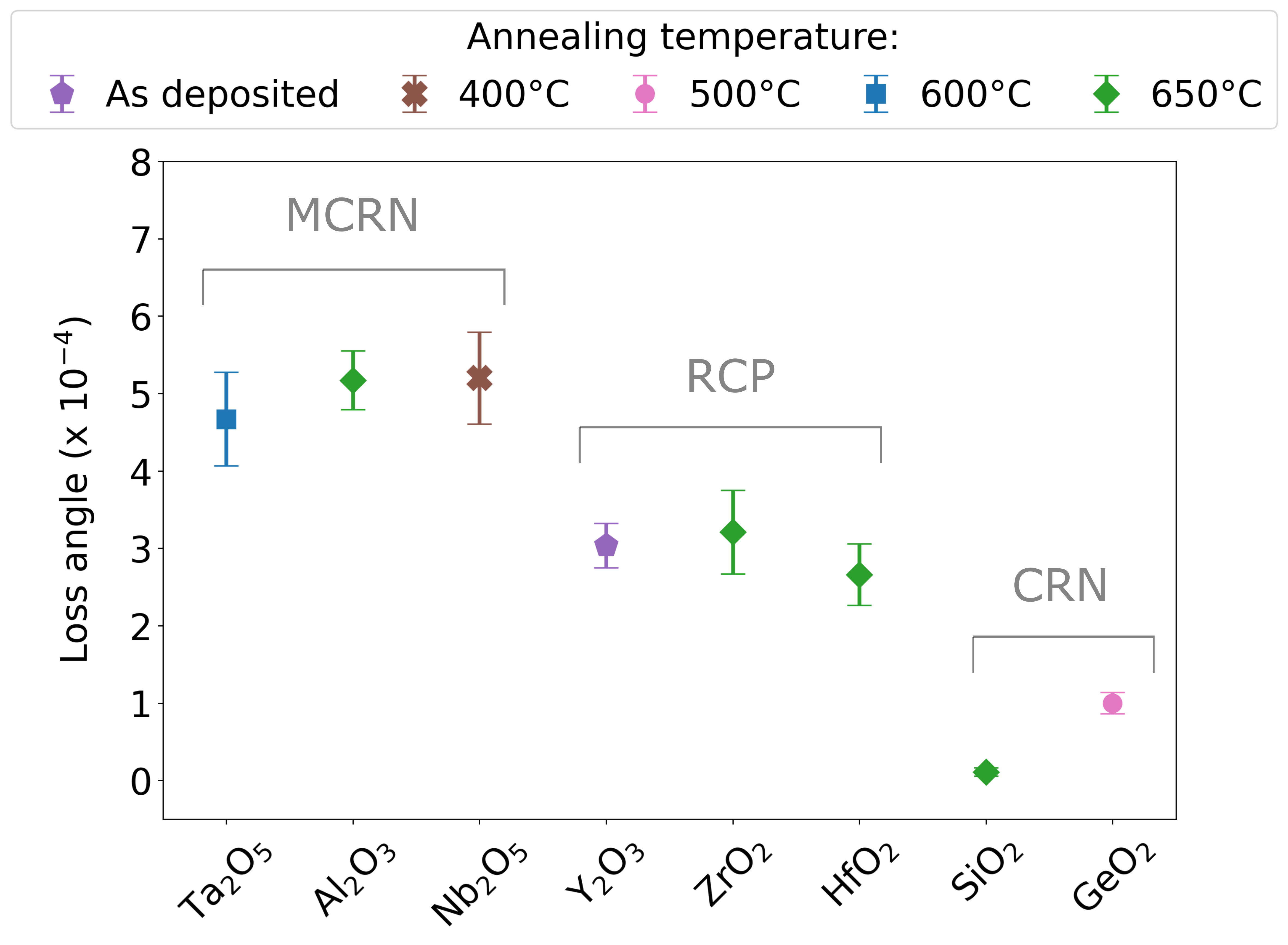}
	\caption{\label{fig:loss_binaries} Coating loss angle at 1 kHz for the evaluated oxide films. Annealing temperatures for each material identify the color of each symbol and are indicated at the top of the figure. The oxides are grouped according to their amorphous morphology classification: modified continuous random network (MCRN), random closed packed (RCP) and continuous random network (CRN).}
\end{figure}

Figure \ref{fig:loss_binaries} presents the loss angle at 1 kHz for the evaluated oxides. Coatings were annealed at the maximum temperature before crystallization is reached, and the corresponding annealing temperatures for each material are shown in the plot. Films that were crystallized as deposited, such as TiO$_2$, ZnO and Sc$_2$O$_3$, were excluded from this study.

Amorphous materials can be classified according to their amorphous morphology as identified by Zallen \cite{zallen2008physics}: continuous random network (CRN), modified continuous random network (MCRN), or random closed packed (RCP). The CRN materials feature mainly covalent bonds with each atom in a bonding state corresponding to its primary chemical valence. For materials with a MCRN morphology, covalent bonds are modified and disrupted by ionic bonds and the average bonding coordination of the oxygen atoms tends to be higher than for CRN materials. Lastly, the RCP morphology consists of a random (non-periodic) closed packing of ions.

Applying this classification to the evaluated oxides, it can be observed that there is a correlation between the amorphous morphology of the material and the mechanical loss as shown in figure \ref{fig:loss_binaries}. The oxides which correspond to a MCRN morphology have the highest room temperature mechanical loss angle values of around $5 \times 10^{-4}$. This group is comprised of Ta$_2$O$_5$, Nb$_2$O$_5$ and Al$_2$O$_3$. The first two are high refractive index materials ($n > 1.50$) at $\lambda$ = 1064 nm and their maximum annealing temperatures are 600\C{} for Ta$_2$O$_5$ and 400\C{} for Nb$_2$O$_5$. On the other hand, Al$_2$O$_3$ has a lower refractive index of around 1.61 and remains amorphous up to 650\C{}.

In the intermediate range of coating loss angle values, around $3 \times 10^{-4}$, the oxides classified as RCP can be found. This group includes the high refractive index materials Y$_2$O$_3$, ZrO$_2$ and HfO$_2$. Y$_2$O$_3$ was only evaluated as deposited given that it crystallized after annealing at 300\C{} leading to increased loss angle values. For ZrO$_2$ and HfO$_2$, annealing temperatures of up to 650\C{} could be applied reaching similar loss angle values to as-deposited Y$_2$O$_3$.

Lastly, the lowest loss angle values are found to correspond to the CRN oxides, SiO$_2$ and GeO$_2$. For SiO$_2$, the loss angle falls below the detection limit of the measurement technique so the value presented corresponds to a lower annealing temperature of 650\C{} for which the loss value could be evaluated. This material remains amorphous up to an annealing temperature of 900\C{}, which also corresponds to the maximum annealing temperature that the silica substrate can withstand without degradation of its optical properties due to impending crystallization of the material. In the case of GeO$_2$, the crystallization temperature is approximately 600\C{} and the loss angle is around $1 \times 10^{-4}$ after annealing at 500\C{} \cite{yang2021enhanced}. Both of these materials have relatively low refractive index, with SiO$_2$ being the current low index material employed in Advanced LIGO and Advanced Virgo.

The Young's modulus and Poisson ratio of the coatings were also determined and are presented in Table \ref{table:mech_prop-oxides}. As mentioned before, SiO$_2$ is the current low refractive index material in the coatings of the test masses of Advanced LIGO and Advanced Virgo. As can be observed from figure \ref{fig:loss_binaries}, SiO$_2$ is in fact the material with low refractive index that presents the lowest mechanical loss value. Regarding the high index materials, Ta$_2$O$_5$ was employed in the first mirrors of LIGO and Virgo. There are several high index oxides in this study that could be employed in a high reflectivity stack along with SiO$_2$: Nb$_2$O$_5$, ZrO$_2$, Ta$_2$O$_5$ and HfO$_2$. Of all these, ZrO$_2$ and HfO$_2$ have the lowest loss angles but their Young's modulus values are four times higher than that of SiO$_2$. According to expression \ref{eq:thermalnoise}, dissimilar Young's modulus values between substrate (SiO$_2$) and coating lead to increased thermal noise. Therefore a stack composed of SiO$_2$ and either ZrO$_2$ or HfO$_2$ will result in elevated thermal noise. Nb$_2$O$_5$ and Ta$_2$O$_5$, on the other hand, feature both a similar Young's modulus to that of SiO$_2$ and a stack using these materials would have a modulus similar to that of the substrate. However, Nb$_2$O$_5$ has a relatively low crystallization temperature allowing a maximum annealing temperature of only 400\C{}. That makes Ta$_2$O$_5$ the most suitable high index material for a stack with reduced thermal noise among the oxides characterized in this study.

\begin{table*}[!ht]
	\setlength\tabcolsep{5pt}
	\begin{center}
		\begin{tabular}{ccccc}
			\hline
			Material & n at $\lambda$ = 1064 nm  & Young's modulus (GPa) & Poisson's ratio\\
			\hline
			Ta$_2$O$_5$ & 2.12 $\pm$ 0.01 & 115 $\pm$ 2 & 0.26 $\pm$ 0.04\\
			Al$_2$O$_3$ & 1.61 $\pm$ 0.01 & 113 $\pm$ 4 & 0.42 $\pm$ 0.05\\
			Nb$_2$O$_5$ & 2.24 $\pm$ 0.01 & 89 $\pm$ 2 & 0.327 $\pm$ 0.002\\
			Y$_2$O$_3$ & 1.92 $\pm$ 0.01 & 151 $\pm$ 4 & 0.34 $\pm$ 0.4\\
			ZrO$_2$  & 2.19 $\pm$ 0.01 & 230 $\pm$ 10 & 0.41 $\pm$ 0.05\\
			HfO$_2$  & 2.09 $\pm$ 0.01 & 248 $\pm$ 6 & 0.32 $\pm$ 0.04\\
			SiO$_2$  & 1.46 $\pm$ 0.01 & 70 $\pm$ 2 & 0.264 $\pm$ 0.002\\
			GeO$_2$ & 1.60 $\pm$ 0.01 & 48.2 $\pm$ 0.6 & 0.29 $\pm$ 0.03\\
			\hline
		\end{tabular}
	\end{center}
	\caption{Refractive index at $\lambda$ = 1064 nm, Young's modulus and Poisson ratio for the evaluated oxide films. All values correspond to films annealed at the temperatures specified in figure \ref{fig:loss_binaries}.}
	\label{table:mech_prop-oxides}
\end{table*}

\subsection{Ta$_2$O$_5$-based mixed oxide coatings}

\begin{figure}[h!]
	\includegraphics[width=\linewidth]{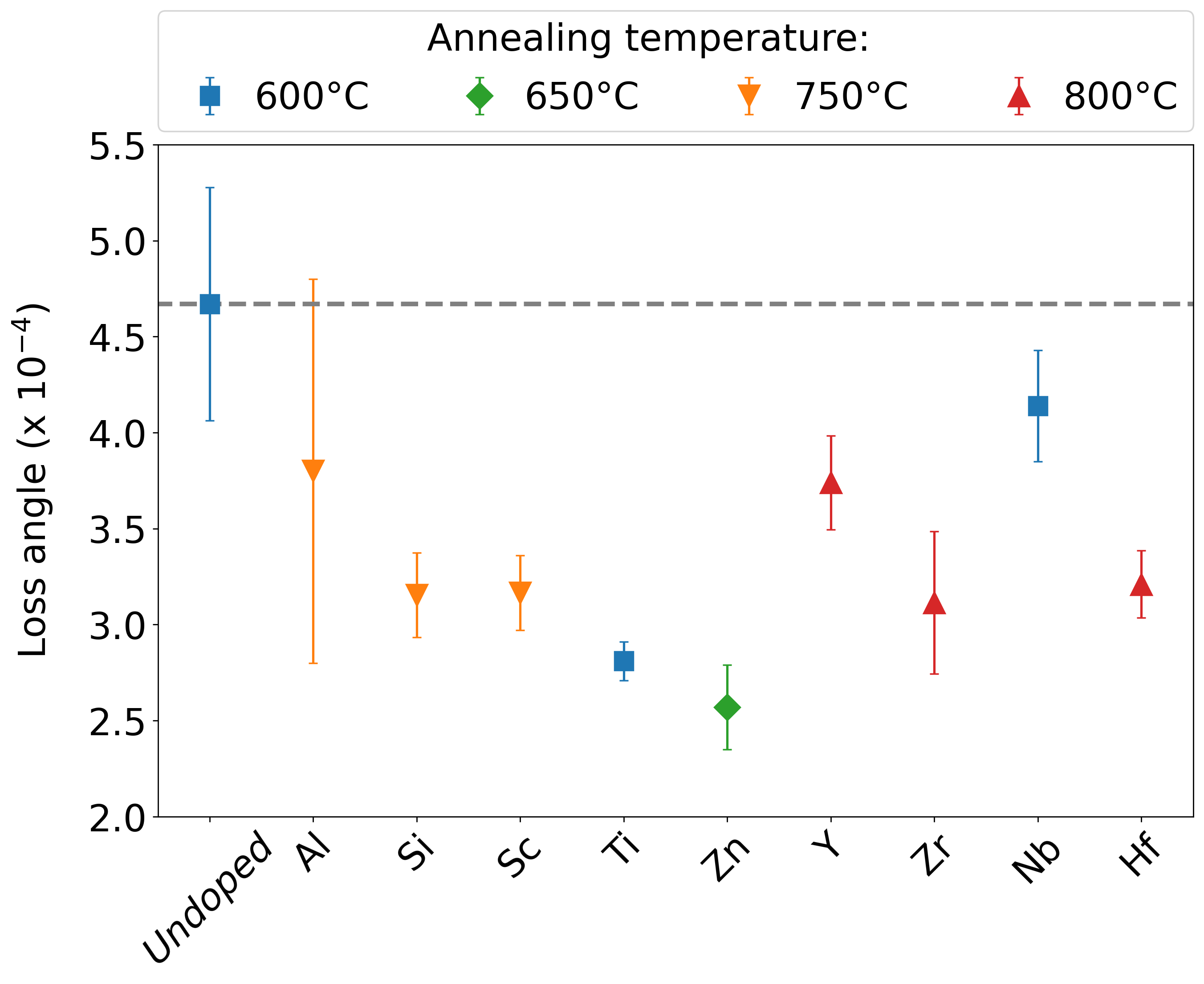}
	\caption{\label{fig:loss_doped} Coating loss angle at 1 kHz for annealed Ta$_2$O$_5$ and Ta$_2$O$_5$-based oxide mixtures. The bottom axis indicates either undoped Ta$_2$O$_5$ or the dopant cation of the mixed film. Annealing temperatures for each material identify the color of each symbol and are indicated at the top of the figure. The grey doted line indicates the loss value for Ta$_2$O$_5$ for ease of comparison with the doped films. Loss angle values for TiO$_2$:Ta$_2$O$_5$ and Ta$_2$O$_5$ are also presented in \cite{fazio2020structure}.}
\end{figure}

Coating loss angle values for Ta$_2$O$_5$ and Ta$_2$O$_5$-based oxide mixtures with cation ratios between 0.1 and 0.3 are shown in figure \ref{fig:loss_doped}. The annealing temperature for each material is also presented, which corresponds to the maximum annealing temperature before crystallization was reached. Comparing with undoped Ta$_2$O$_5$, the addition of the dopant in most cases leads to a decrease in the mechanical loss. Only for Al$_2$O$_3$:Ta$_2$O$_5$ and Nb$_2$O$_5$:Ta$_2$O$_5$ the dopant does not modify the loss significantly. The lowest loss values are obtained for TiO$_2$:Ta$_2$O$_5$ and ZnO:Ta$_2$O$_5$, with the first being the approximate composition of the high refractive index material currently employed in the mirrors of Advanced LIGO and Advanced Virgo \cite{abbott2016gw150914,acernese2014advanced} .

\begin{figure}[h!] 
	\includegraphics[width=\linewidth]{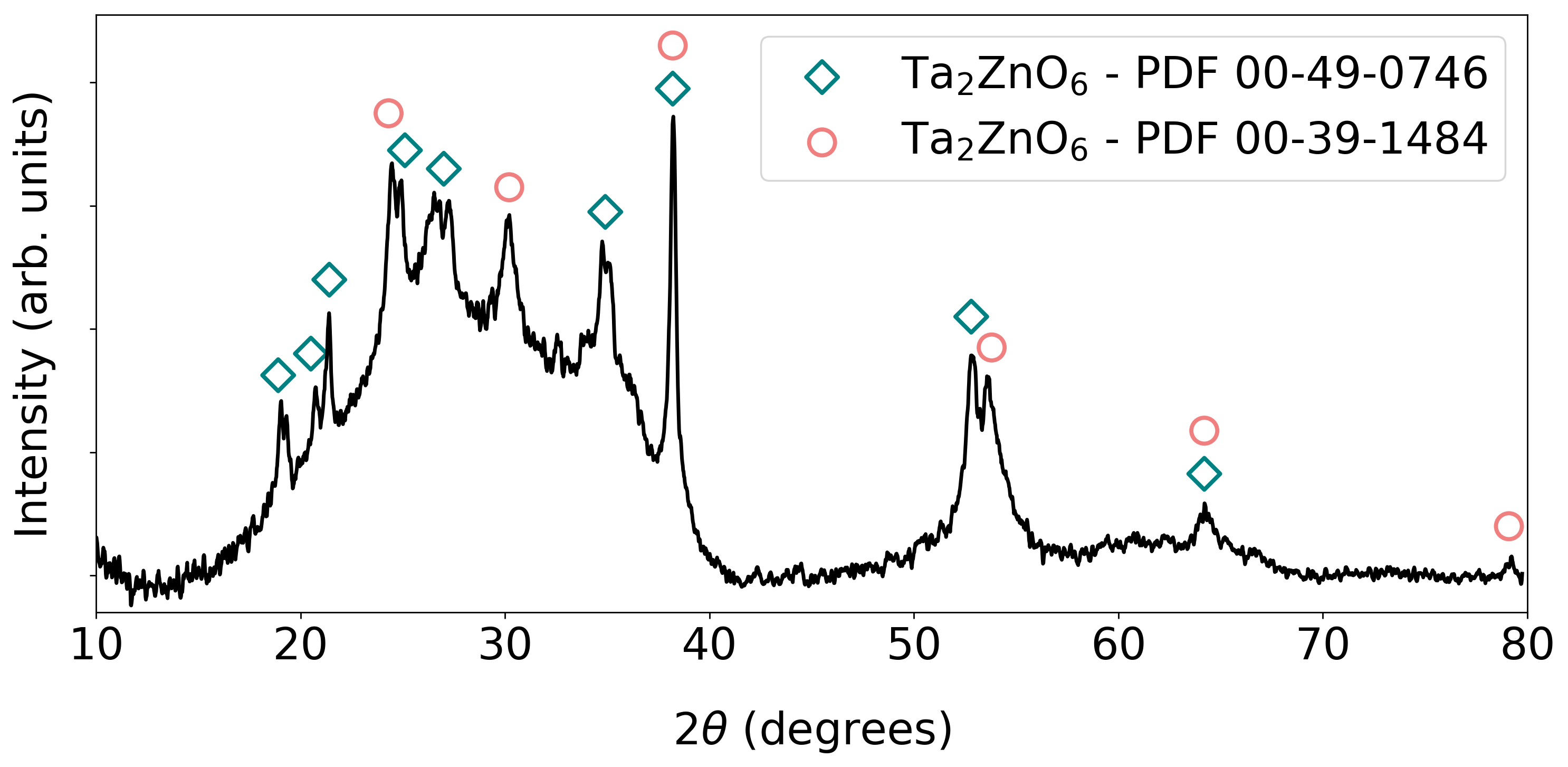}
	\caption{\label{fig:zn_gaxrd} GIXRD diffractogram of ZnO:Ta$_2$O$_5$ after annealing at 650\C{}. Tabulated peak positions for two polymorphs of Ta$_2$ZnO$_6$ (reference pattern PDF 00-049-0746 and 00-39-1484 \cite{gates2019powder}) are included.}
\end{figure}

In most cases the dopant increases the crystallization temperature of Ta$_2$O$_5$, as can be inferred from figure \ref{fig:loss_doped}. Generally, the loss angle of a given material decreases with increasing annealing temperature until crystallization is reached. However, there is no correlation between crystallization temperature and loss angle when comparing among different mixed oxides. In other words, a Ta$_2$O$_5$-based mixture with a high crystallization temperature will not necessarily feature the lowest loss angle as can be observed in figure \ref{fig:loss_doped}. Further analysis of the films carried out in \cite{fazio2021prediction} indicates that the dopant either acts as an amorphizer agent or induces the formation of a ternary phase. Among all the evaluated coatings, a ternary compound was only found to form in the case of TiO$_2$:Ta$_2$O$_5$ and ZnO:Ta$_2$O$_5$ which feature the lowest mechanical losses. Previous research on TiO$_2$:Ta$_2$O$_5$ indicated that the ternary phase formation introduced significant morphological and structural changes that affected the mixed oxide coating and were also linked to the decreased loss value \cite{fazio2020structure}. For ZnO:Ta$_2$O$_5$ the lowest loss angle was achieved after annealing at 650\C{} but, contrary to all other evaluated films in this study (oxides and mixed oxides), the structure was partially crystallized. Figure \ref{fig:zn_gaxrd} shows the corresponding GIXRD diffractogram that features an amorphous background with superimposed peaks that correspond to two polymorphs of the ternary compound Ta$_2$ZnO$_6$ (reference pattern PDF 00-049-0746 and 00-39-1484 \cite{gates2019powder}). It is notable that the presence of a crystalline phase does not lead to higher loss values, as the grain boundaries are known to increase the loss significantly. Generally, the mechanical loss angle is particularly sensitive to even partial crystallization which causes a dramatic increase in the loss.

Table \ref{table:mech_prop-doped} presents the refractive index at $\lambda$ = 1064 nm, Young's modulus and Poisson ratio for the Ta$_2$O$_5$-based mixtures. All values correspond to films annealed at their maximum annealing temperature, specified in figure \ref{fig:loss_doped}. The refractive index values roughly follow the law of mixtures (that is, a weighted average), with dopants of low index such as Al$_2$O$_3$ leading to a mixed film with lower refractive index than Ta$_2$O$_5$. In the case of Sc$_2$O$_3$:Ta$_2$O$_5$ a more complex relation was found, given that the dopant induces the presence of a tantalum suboxide compound as described in \cite{fazio2020growth}. For TiO$_2$:Ta$_2$O$_5$, the annealing promotes the formation of Ar-filled bubbles in the coatings which affect the refractive index as observed in \cite{fazio2020structure}. Overall, most mixed films have a lower refractive index than Ta$_2$O$_5$, with the index of SiO$_2$:Ta$_2$O$_5$ being the lowest of the evaluated mixtures. The Young's modulus, an important factor that contributes to the Brownian thermal noise, shows variations of up to 27\% compared with undoped Ta$_2$O$_5$. This small variation in the Young's modulus compared to Ta$_2$O$_5$ is due to the low cation ratio of these mixtures and the specific Young's modulus of these materials and consequently should have little to no effect in the Brownian noise of an HR stack designed using a Ta$_2$O$_5$-based mixed oxide as the high index material.

\begin{table*}[!ht]
	\setlength\tabcolsep{5pt}
	\begin{center}
		\begin{tabular}{ccccc}
			\hline
			Material & n at $\lambda$ = 1064 nm & Young's modulus (GPa) & Poisson's ratio\\
			\hline
			Al$_2$O$_3$:Ta$_2$O$_5$ & 2.01 $\pm$ 0.01 & 132 $\pm$ 6 & 0.37 $\pm$ 0.08\\
			SiO$_2$:Ta$_2$O$_5$ & 1.93 $\pm$ 0.01 & 121 $\pm$ 2 & 0.39 $\pm$ 0.02\\
			Sc$_2$O$_3$:Ta$_2$O$_5$ & 2.09 $\pm$ 0.01 & 133 $\pm$ 2 & 0.39 $\pm$ 0.03\\
			TiO$_2$:Ta$_2$O$_5$ & 2.19 $\pm$ 0.01 & 128 $\pm$ 4 & 0.35 $\pm$ 0.02\\
			ZnO:Ta$_2$O$_5$ & 2.05 $\pm$ 0.01 & 103 $\pm$ 2 & 0.49 $\pm$ 0.02\\
			Y$_2$O$_3$:Ta$_2$O$_5$ & 2.03 $\pm$ 0.01 & 123 $\pm$ 5 & 0.35 $\pm$ 0.07\\
			ZrO$_2$:Ta$_2$O$_5$  & 2.07 $\pm$ 0.01 & 143 $\pm$ 5 & 0.36 $\pm$ 0.05\\
			Nb$_2$O$_5$:Ta$_2$O$_5$ & 2.11 $\pm$ 0.01 & 118 $\pm$ 4 & 0.3 $\pm$ 0.1\\
			HfO$_2$:Ta$_2$O$_5$  & 2.05 $\pm$ 0.01 & 146 $\pm$ 5 & 0.32 $\pm$ 0.06\\			
			\hline
		\end{tabular}
	\end{center}
	\caption{Refractive index at $\lambda$ = 1064 nm, Young's modulus and Poisson ratio for the evaluated Ta$_2$O$_5$-based mixed oxide films. All values correspond to films annealed at the temperatures specified in figure \ref{fig:loss_doped}. Data corresponding to TiO$_2$:Ta$_2$O$_5$ are also presented in \cite{fazio2020structure, vajente2020}.}
	\label{table:mech_prop-doped}
\end{table*}

\begin{figure}[h!]
	\includegraphics[width=\linewidth]{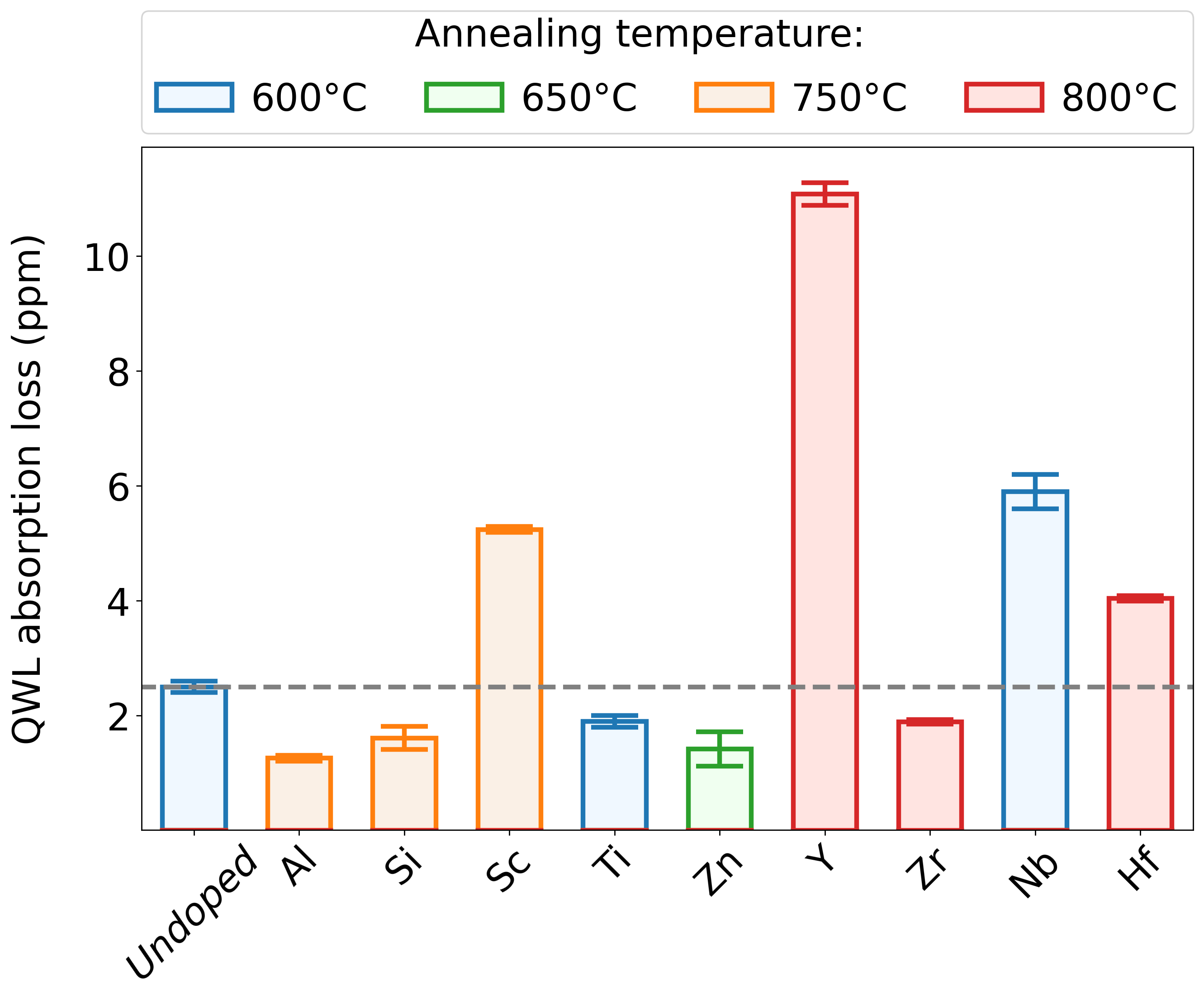}
	\caption{\label{fig:absorption_doped} Absorption loss at $\lambda$ = 1064 nm for Ta$_2$O$_5$ and Ta$_2$O$_5$-based mixed oxide films normalized to quarter wavelength optical thickness. The bottom axis indicates either undoped Ta$_2$O$_5$ or the dopant cation of the mixed film. The grey doted line indicates the absorption loss value for Ta$_2$O$_5$ for ease of comparison with the doped films. Data corresponding to Ta$_2$O$_5$, TiO$_2$:Ta$_2$O$_5$ and Sc$_2$O$_3$:Ta$_2$O$_5$ are also presented in \cite{fazio2020growth,fazio2020structure}}
\end{figure}

The absorption loss at $\lambda$ = 1064 nm measured for annealed Ta$_2$O$_5$ and Ta$_2$O$_5$-based mixtures can be observed in figure \ref{fig:absorption_doped}. The values are normalized to a quarter wavelength (QWL) optical thickness, as typical HR stacks will be composed of QWL layers designed for maximum reflectivity at the laser wavelength of $\lambda$ = 1064 nm. It can be seen that the dopant can increase or decrease the absorption loss compared to undoped Ta$_2$O$_5$. The mixtures with the lowest absorption loss values are Al$_2$O$_3$:Ta$_2$O$_5$, SiO$_2$:Ta$_2$O$_5$, TiO$_2$:Ta$_2$O$_5$ ZnO:Ta$_2$O$_5$ and ZrO$_2$:Ta$_2$O$_5$. On the other hand, Sc$_2$O$_3$:Ta$_2$O$_5$, Y$_2$O$_3$:Ta$_2$O$_5$, Nb$_2$O$_5$:Ta$_2$O$_5$ and HfO$_2$:Ta$_2$O$_5$ all have absorption loss values much larger than undoped Ta$_2$O$_5$.

\begin{figure}[h!] 
	\includegraphics[width=0.98\linewidth]{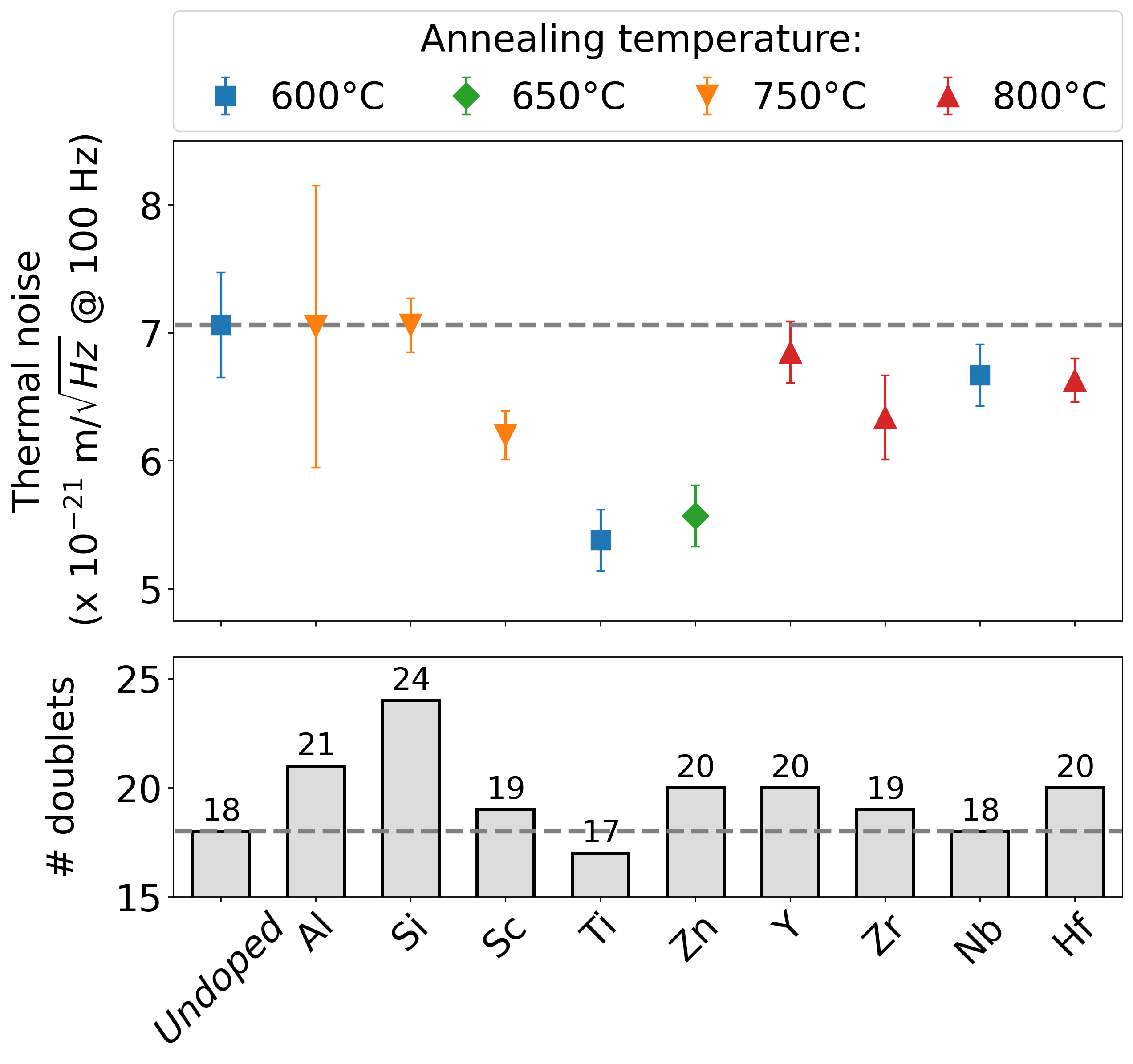}
	\caption{\label{fig:noise_doped} Top panel: comparison of the Brownian thermal noise level attainable with Ta$_2$O$_5$ and Ta$_2$O$_5$-based mixed oxide materials. Annealing temperatures are specified for each material. Bottom panel: number of doublets used in each HR stack design to reach a reflectivity of around 5 ppm. The bottom axis indicates either undoped Ta$_2$O$_5$ or the dopant cation of the mixed film. In both panels, the grey dashed lines indicate the values for Ta$_2$O$_5$ for ease of comparison with the doped films.}
\end{figure}

In order to fully evaluate the suitability of the Ta$_2$O$_5$-based mixed oxide coatings as high index components in an HR stack, the Brownian thermal noise was calculated for stacks composed of layers of SiO$_2$ and each mixture material. Figure \ref{fig:noise_doped} presents the Brownian noise level in the top panel and the number of doublets (pair of SiO$_2$ and high index layer of QWL optical thickness each) used in the stack design to reach a 5 ppm transmissivity. As expected, the total number of layers is inversely proportional to the refractive index contrast between the mixed oxide material and SiO$_2$ with SiO$_2$:Ta$_2$O$_5$ requiring the largest number of doublets in the design.

The Brownian thermal noise takes into account not only the loss angle of the material but also the total thickness of the high and low index materials that compose the HR stack. The ideal high index material would be one that features the lowest mechanical loss and highest refractive index at $\lambda$ = 1064 nm. However, a decrease in the mechanical loss angle can be offset by a decrease in the refractive index of the material which leads to more doublets being required in the stack to achieve the desired reflectivity. A clear example of this is the case of SiO$_2$:Ta$_2$O$_5$ for which the mechanical loss angle was significantly lower than undoped Ta$_2$O$_5$ but so was the refractive index, and therefore the resulting Brownian noise is of similar value. The use of dopants that decrease the refractive index can be challenging, as only a significant decrease in the mechanical loss can provide a lower Brownian noise. Most dopants lowered the mechanical loss angle but only a few of them contributed to lowering the Brownian noise. The most viable candidates of all the evaluated mixtures are Sc$_2$O$_3$:Ta$_2$O$_5$, TiO$_2$:Ta$_2$O$_5$ and ZnO:Ta$_2$O$_5$. The first features elevated absorption loss and the presence of tantalum suboxide compounds, while the later is partially crystallized and thus may increase the bulk scattering losses. Our results confirm, within the tested mixed oxide materials, that the only suitable dopant for Ta$_2$O$_5$ with low cation concentration appears to be TiO$_2$ which is the mixture already in use in Advanced LIGO and Advanced Virgo.

\section{Conclusions} \label{sec:conclusions}

A systematic study of amorphous metal oxide and Ta$_2$O$_5$-based mixed oxide coatings was carried out to evaluate their suitability for low Brownian noise interference coatings for GW detectors. Coatings were characterized to obtain their optical, structural and elastic properties, in addition to measuring the loss angle of the materials. The metal oxides evaluated in this study allowed to identify a correlation between the mechanical loss and the amorphous morphology of the material. In particular, continuous random network oxides such as SiO$_2$ and GeO$_2$ featured the lowest loss angles. A comprehensive survey of Ta$_2$O$_5$-based mixed oxide thin films provided insights into the effect of the dopant in the elastic and optical properties of the coatings as well as in the mechanical loss. We estimated the thermal noise associated with high-reflectance multilayer stacks that employ each of the mixed oxides as the high index material. We concluded that the current high index material of TiO$_2$-doped Ta$_2$O$_5$ is the optimal choice for reduced thermal noise among Ta$_2$O$_5$-based mixed oxide coatings with low dopant concentrations. With the upcoming A+ Ligo upgrade and the advent of third generation GW detectors with stricter noise budgets, a comprehensive understanding of the properties of amorphous materials is critical for the design of future interference coatings optimized for low Brownian noise. The results of the present study could aid in the development of novel materials optimized for use in GW detectors, leveraging off the relative ease of production of metal oxides and mixed metal oxides that fulfill the strict requirements of this application.

\begin{acknowledgments}

This work was supported by the National Science Foundation - Moore Foundation Center for Coatings Research under NSF award No. 1708010 and NSF award No. 1710957.

\end{acknowledgments}

\bibliography{biblio}

\begin{thebibliography}{35}%
\makeatletter
\providecommand \@ifxundefined [1]{%
 \@ifx{#1\undefined}
}%
\providecommand \@ifnum [1]{%
 \ifnum #1\expandafter \@firstoftwo
 \else \expandafter \@secondoftwo
 \fi
}%
\providecommand \@ifx [1]{%
 \ifx #1\expandafter \@firstoftwo
 \else \expandafter \@secondoftwo
 \fi
}%
\providecommand \natexlab [1]{#1}%
\providecommand \enquote  [1]{``#1''}%
\providecommand \bibnamefont  [1]{#1}%
\providecommand \bibfnamefont [1]{#1}%
\providecommand \citenamefont [1]{#1}%
\providecommand \href@noop [0]{\@secondoftwo}%
\providecommand \href [0]{\begingroup \@sanitize@url \@href}%
\providecommand \@href[1]{\@@startlink{#1}\@@href}%
\providecommand \@@href[1]{\endgroup#1\@@endlink}%
\providecommand \@sanitize@url [0]{\catcode `\\12\catcode `\$12\catcode
  `\&12\catcode `\#12\catcode `\^12\catcode `\_12\catcode `\%12\relax}%
\providecommand \@@startlink[1]{}%
\providecommand \@@endlink[0]{}%
\providecommand \url  [0]{\begingroup\@sanitize@url \@url }%
\providecommand \@url [1]{\endgroup\@href {#1}{\urlprefix }}%
\providecommand \urlprefix  [0]{URL }%
\providecommand \Eprint [0]{\href }%
\providecommand \doibase [0]{https://doi.org/}%
\providecommand \selectlanguage [0]{\@gobble}%
\providecommand \bibinfo  [0]{\@secondoftwo}%
\providecommand \bibfield  [0]{\@secondoftwo}%
\providecommand \translation [1]{[#1]}%
\providecommand \BibitemOpen [0]{}%
\providecommand \bibitemStop [0]{}%
\providecommand \bibitemNoStop [0]{.\EOS\space}%
\providecommand \EOS [0]{\spacefactor3000\relax}%
\providecommand \BibitemShut  [1]{\csname bibitem#1\endcsname}%
\let\auto@bib@innerbib\@empty
\bibitem [{\citenamefont {Abbott}\ \emph {et~al.}(2016)\citenamefont {Abbott},
  \citenamefont {Abbott}, \citenamefont {Abbott}, \citenamefont {Abernathy},
  \citenamefont {Acernese}, \citenamefont {Ackley}, \citenamefont {Adams},
  \citenamefont {Adams}, \citenamefont {Addesso}, \citenamefont {Adhikari}
  \emph {et~al.}}]{abbott2016gw150914}%
  \BibitemOpen
  \bibfield  {author} {\bibinfo {author} {\bibfnamefont {B.~P.}\ \bibnamefont
  {Abbott}}, \bibinfo {author} {\bibfnamefont {R.}~\bibnamefont {Abbott}},
  \bibinfo {author} {\bibfnamefont {T.}~\bibnamefont {Abbott}}, \bibinfo
  {author} {\bibfnamefont {M.}~\bibnamefont {Abernathy}}, \bibinfo {author}
  {\bibfnamefont {F.}~\bibnamefont {Acernese}}, \bibinfo {author}
  {\bibfnamefont {K.}~\bibnamefont {Ackley}}, \bibinfo {author} {\bibfnamefont
  {C.}~\bibnamefont {Adams}}, \bibinfo {author} {\bibfnamefont
  {T.}~\bibnamefont {Adams}}, \bibinfo {author} {\bibfnamefont
  {P.}~\bibnamefont {Addesso}}, \bibinfo {author} {\bibfnamefont
  {R.}~\bibnamefont {Adhikari}}, \emph {et~al.},\ }\bibfield  {title} {\bibinfo
  {title} {Gw150914: The advanced ligo detectors in the era of first
  discoveries},\ }\href@noop {} {\bibfield  {journal} {\bibinfo  {journal}
  {Physical Review Letters}\ }\textbf {\bibinfo {volume} {116}},\ \bibinfo
  {pages} {131103} (\bibinfo {year} {2016})}\BibitemShut {NoStop}%
\bibitem [{\citenamefont {Acernese}\ \emph {et~al.}(2014)\citenamefont
  {Acernese}, \citenamefont {Agathos}, \citenamefont {Agatsuma}, \citenamefont
  {Aisa}, \citenamefont {Allemandou}, \citenamefont {Allocca}, \citenamefont
  {Amarni}, \citenamefont {Astone}, \citenamefont {Balestri}, \citenamefont
  {Ballardin} \emph {et~al.}}]{acernese2014advanced}%
  \BibitemOpen
  \bibfield  {author} {\bibinfo {author} {\bibfnamefont {F.}~\bibnamefont
  {Acernese}}, \bibinfo {author} {\bibfnamefont {M.}~\bibnamefont {Agathos}},
  \bibinfo {author} {\bibfnamefont {K.}~\bibnamefont {Agatsuma}}, \bibinfo
  {author} {\bibfnamefont {D.}~\bibnamefont {Aisa}}, \bibinfo {author}
  {\bibfnamefont {N.}~\bibnamefont {Allemandou}}, \bibinfo {author}
  {\bibfnamefont {A.}~\bibnamefont {Allocca}}, \bibinfo {author} {\bibfnamefont
  {J.}~\bibnamefont {Amarni}}, \bibinfo {author} {\bibfnamefont
  {P.}~\bibnamefont {Astone}}, \bibinfo {author} {\bibfnamefont
  {G.}~\bibnamefont {Balestri}}, \bibinfo {author} {\bibfnamefont
  {G.}~\bibnamefont {Ballardin}}, \emph {et~al.},\ }\bibfield  {title}
  {\bibinfo {title} {Advanced virgo: a second-generation interferometric
  gravitational wave detector},\ }\href@noop {} {\bibfield  {journal} {\bibinfo
   {journal} {Classical and Quantum Gravity}\ }\textbf {\bibinfo {volume}
  {32}},\ \bibinfo {pages} {024001} (\bibinfo {year} {2014})}\BibitemShut
  {NoStop}%
\bibitem [{\citenamefont {Akutsu}\ \emph {et~al.}(2015)\citenamefont {Akutsu},
  \citenamefont {collaboration} \emph {et~al.}}]{akutsu2015large}%
  \BibitemOpen
  \bibfield  {author} {\bibinfo {author} {\bibfnamefont {T.}~\bibnamefont
  {Akutsu}}, \bibinfo {author} {\bibfnamefont {K.}~\bibnamefont
  {collaboration}}, \emph {et~al.},\ }\bibfield  {title} {\bibinfo {title}
  {{Large-scale cryogenic gravitational-wave telescope in Japan: KAGRA}},\ }in\
  \href@noop {} {\emph {\bibinfo {booktitle} {Journal of Physics: Conference
  Series}}},\ Vol.\ \bibinfo {volume} {610}\ (\bibinfo {organization} {IOP
  Publishing},\ \bibinfo {year} {2015})\ p.\ \bibinfo {pages}
  {012016}\BibitemShut {NoStop}%
\bibitem [{\citenamefont {Buonanno}\ and\ \citenamefont
  {Chen}(2001)}]{buonanno2001quantum}%
  \BibitemOpen
  \bibfield  {author} {\bibinfo {author} {\bibfnamefont {A.}~\bibnamefont
  {Buonanno}}\ and\ \bibinfo {author} {\bibfnamefont {Y.}~\bibnamefont
  {Chen}},\ }\bibfield  {title} {\bibinfo {title} {Quantum noise in second
  generation, signal-recycled laser interferometric gravitational-wave
  detectors},\ }\href@noop {} {\bibfield  {journal} {\bibinfo  {journal}
  {Physical Review D}\ }\textbf {\bibinfo {volume} {64}},\ \bibinfo {pages}
  {042006} (\bibinfo {year} {2001})}\BibitemShut {NoStop}%
\bibitem [{\citenamefont {Saulson}(1990)}]{saulson1990thermal}%
  \BibitemOpen
  \bibfield  {author} {\bibinfo {author} {\bibfnamefont {P.~R.}\ \bibnamefont
  {Saulson}},\ }\bibfield  {title} {\bibinfo {title} {Thermal noise in
  mechanical experiments},\ }\href@noop {} {\bibfield  {journal} {\bibinfo
  {journal} {Physical Review D}\ }\textbf {\bibinfo {volume} {42}},\ \bibinfo
  {pages} {2437} (\bibinfo {year} {1990})}\BibitemShut {NoStop}%
\bibitem [{\citenamefont {Numata}\ \emph {et~al.}(2004)\citenamefont {Numata},
  \citenamefont {Kemery},\ and\ \citenamefont {Camp}}]{numata2004thermal}%
  \BibitemOpen
  \bibfield  {author} {\bibinfo {author} {\bibfnamefont {K.}~\bibnamefont
  {Numata}}, \bibinfo {author} {\bibfnamefont {A.}~\bibnamefont {Kemery}},\
  and\ \bibinfo {author} {\bibfnamefont {J.}~\bibnamefont {Camp}},\ }\bibfield
  {title} {\bibinfo {title} {Thermal-noise limit in the frequency stabilization
  of lasers with rigid cavities},\ }\href@noop {} {\bibfield  {journal}
  {\bibinfo  {journal} {Physical review letters}\ }\textbf {\bibinfo {volume}
  {93}},\ \bibinfo {pages} {250602} (\bibinfo {year} {2004})}\BibitemShut
  {NoStop}%
\bibitem [{\citenamefont {Ludlow}\ \emph {et~al.}(2008)\citenamefont {Ludlow},
  \citenamefont {Zelevinsky}, \citenamefont {Campbell}, \citenamefont {Blatt},
  \citenamefont {Boyd}, \citenamefont {de~Miranda}, \citenamefont {Martin},
  \citenamefont {Thomsen}, \citenamefont {Foreman}, \citenamefont {Ye} \emph
  {et~al.}}]{ludlow2008sr}%
  \BibitemOpen
  \bibfield  {author} {\bibinfo {author} {\bibfnamefont {A.~D.}\ \bibnamefont
  {Ludlow}}, \bibinfo {author} {\bibfnamefont {T.}~\bibnamefont {Zelevinsky}},
  \bibinfo {author} {\bibfnamefont {G.}~\bibnamefont {Campbell}}, \bibinfo
  {author} {\bibfnamefont {S.}~\bibnamefont {Blatt}}, \bibinfo {author}
  {\bibfnamefont {M.}~\bibnamefont {Boyd}}, \bibinfo {author} {\bibfnamefont
  {M.~H.}\ \bibnamefont {de~Miranda}}, \bibinfo {author} {\bibfnamefont
  {M.}~\bibnamefont {Martin}}, \bibinfo {author} {\bibfnamefont
  {J.}~\bibnamefont {Thomsen}}, \bibinfo {author} {\bibfnamefont {S.~M.}\
  \bibnamefont {Foreman}}, \bibinfo {author} {\bibfnamefont {J.}~\bibnamefont
  {Ye}}, \emph {et~al.},\ }\bibfield  {title} {\bibinfo {title} {{Sr lattice
  clock at 1$\times$ 10--16 fractional uncertainty by remote optical evaluation
  with a Ca clock}},\ }\href@noop {} {\bibfield  {journal} {\bibinfo  {journal}
  {Science}\ }\textbf {\bibinfo {volume} {319}},\ \bibinfo {pages} {1805}
  (\bibinfo {year} {2008})}\BibitemShut {NoStop}%
\bibitem [{\citenamefont {Rosenband}\ \emph {et~al.}(2008)\citenamefont
  {Rosenband}, \citenamefont {Hume}, \citenamefont {Schmidt}, \citenamefont
  {Chou}, \citenamefont {Brusch}, \citenamefont {Lorini}, \citenamefont
  {Oskay}, \citenamefont {Drullinger}, \citenamefont {Fortier}, \citenamefont
  {Stalnaker} \emph {et~al.}}]{rosenband2008frequency}%
  \BibitemOpen
  \bibfield  {author} {\bibinfo {author} {\bibfnamefont {T.}~\bibnamefont
  {Rosenband}}, \bibinfo {author} {\bibfnamefont {D.}~\bibnamefont {Hume}},
  \bibinfo {author} {\bibfnamefont {P.}~\bibnamefont {Schmidt}}, \bibinfo
  {author} {\bibfnamefont {C.-W.}\ \bibnamefont {Chou}}, \bibinfo {author}
  {\bibfnamefont {A.}~\bibnamefont {Brusch}}, \bibinfo {author} {\bibfnamefont
  {L.}~\bibnamefont {Lorini}}, \bibinfo {author} {\bibfnamefont
  {W.}~\bibnamefont {Oskay}}, \bibinfo {author} {\bibfnamefont {R.~E.}\
  \bibnamefont {Drullinger}}, \bibinfo {author} {\bibfnamefont {T.~M.}\
  \bibnamefont {Fortier}}, \bibinfo {author} {\bibfnamefont {J.~E.}\
  \bibnamefont {Stalnaker}}, \emph {et~al.},\ }\bibfield  {title} {\bibinfo
  {title} {{Frequency ratio of Al+ and Hg+ single-ion optical clocks; metrology
  at the 17th decimal place}},\ }\href@noop {} {\bibfield  {journal} {\bibinfo
  {journal} {Science}\ }\textbf {\bibinfo {volume} {319}},\ \bibinfo {pages}
  {1808} (\bibinfo {year} {2008})}\BibitemShut {NoStop}%
\bibitem [{\citenamefont {Bishof}\ \emph {et~al.}(2013)\citenamefont {Bishof},
  \citenamefont {Zhang}, \citenamefont {Martin},\ and\ \citenamefont
  {Ye}}]{bishof2013optical}%
  \BibitemOpen
  \bibfield  {author} {\bibinfo {author} {\bibfnamefont {M.}~\bibnamefont
  {Bishof}}, \bibinfo {author} {\bibfnamefont {X.}~\bibnamefont {Zhang}},
  \bibinfo {author} {\bibfnamefont {M.~J.}\ \bibnamefont {Martin}},\ and\
  \bibinfo {author} {\bibfnamefont {J.}~\bibnamefont {Ye}},\ }\bibfield
  {title} {\bibinfo {title} {Optical spectrum analyzer with quantum-limited
  noise floor},\ }\href@noop {} {\bibfield  {journal} {\bibinfo  {journal}
  {Physical Review Letters}\ }\textbf {\bibinfo {volume} {111}},\ \bibinfo
  {pages} {093604} (\bibinfo {year} {2013})}\BibitemShut {NoStop}%
\bibitem [{\citenamefont {Granata}\ \emph
  {et~al.}(2020{\natexlab{a}})\citenamefont {Granata}, \citenamefont {Amato},
  \citenamefont {Balzarini}, \citenamefont {Canepa}, \citenamefont {Degallaix},
  \citenamefont {Forest}, \citenamefont {Dolique}, \citenamefont {Mereni},
  \citenamefont {Michel}, \citenamefont {Pinard} \emph
  {et~al.}}]{granata2020amorphous}%
  \BibitemOpen
  \bibfield  {author} {\bibinfo {author} {\bibfnamefont {M.}~\bibnamefont
  {Granata}}, \bibinfo {author} {\bibfnamefont {A.}~\bibnamefont {Amato}},
  \bibinfo {author} {\bibfnamefont {L.}~\bibnamefont {Balzarini}}, \bibinfo
  {author} {\bibfnamefont {M.}~\bibnamefont {Canepa}}, \bibinfo {author}
  {\bibfnamefont {J.}~\bibnamefont {Degallaix}}, \bibinfo {author}
  {\bibfnamefont {D.}~\bibnamefont {Forest}}, \bibinfo {author} {\bibfnamefont
  {V.}~\bibnamefont {Dolique}}, \bibinfo {author} {\bibfnamefont
  {L.}~\bibnamefont {Mereni}}, \bibinfo {author} {\bibfnamefont
  {C.}~\bibnamefont {Michel}}, \bibinfo {author} {\bibfnamefont
  {L.}~\bibnamefont {Pinard}}, \emph {et~al.},\ }\bibfield  {title} {\bibinfo
  {title} {Amorphous optical coatings of present gravitational-wave
  interferometers},\ }\href@noop {} {\bibfield  {journal} {\bibinfo  {journal}
  {Classical and Quantum Gravity}\ }\textbf {\bibinfo {volume} {37}},\ \bibinfo
  {pages} {095004} (\bibinfo {year} {2020}{\natexlab{a}})}\BibitemShut
  {NoStop}%
\bibitem [{\citenamefont {Martin}\ and\ \citenamefont
  {Reid}(2012)}]{martin2012coating}%
  \BibitemOpen
  \bibfield  {author} {\bibinfo {author} {\bibfnamefont {I.}~\bibnamefont
  {Martin}}\ and\ \bibinfo {author} {\bibfnamefont {S.}~\bibnamefont {Reid}},\
  }{\selectlanguage {English}\bibinfo {title} {Coating thermal noise}},\ in\
  \href {https://doi.org/10.1017/CBO9780511762314.006} {{\selectlanguage
  {English}\emph {\bibinfo {booktitle} {Optical Coatings and Thermal Noise in
  Precision Measurement}}}},\ \bibinfo {editor} {edited by\ \bibinfo {editor}
  {\bibfnamefont {G.}~\bibnamefont {Harry}}, \bibinfo {editor} {\bibfnamefont
  {T.}~\bibnamefont {Bodiya}},\ and\ \bibinfo {editor} {\bibfnamefont
  {R.}~\bibnamefont {DeSalvo}}}\ (\bibinfo  {publisher} {Cambridge University
  Press},\ \bibinfo {address} {United Kingdom},\ \bibinfo {year} {2012})\ pp.\
  \bibinfo {pages} {31--54}\BibitemShut {NoStop}%
\bibitem [{\citenamefont {Fejer}(2021)}]{fejer2021effective}%
  \BibitemOpen
  \bibfield  {author} {\bibinfo {author} {\bibfnamefont {M.~M.}\ \bibnamefont
  {Fejer}},\ }\bibfield  {title} {\bibinfo {title} {Effective medium
  description of multilayer coatings},\ }\href@noop {} {\bibfield  {journal}
  {\bibinfo  {journal} {LIGO Document}\ } (\bibinfo {year} {2021})}\BibitemShut
  {NoStop}%
\bibitem [{\citenamefont {Penn}\ \emph {et~al.}(2003)\citenamefont {Penn},
  \citenamefont {Sneddon}, \citenamefont {Armandula}, \citenamefont
  {Betzwieser}, \citenamefont {Cagnoli}, \citenamefont {Camp}, \citenamefont
  {Crooks}, \citenamefont {Fejer}, \citenamefont {Gretarsson}, \citenamefont
  {Harry} \emph {et~al.}}]{penn2003mechanical}%
  \BibitemOpen
  \bibfield  {author} {\bibinfo {author} {\bibfnamefont {S.~D.}\ \bibnamefont
  {Penn}}, \bibinfo {author} {\bibfnamefont {P.~H.}\ \bibnamefont {Sneddon}},
  \bibinfo {author} {\bibfnamefont {H.}~\bibnamefont {Armandula}}, \bibinfo
  {author} {\bibfnamefont {J.~C.}\ \bibnamefont {Betzwieser}}, \bibinfo
  {author} {\bibfnamefont {G.}~\bibnamefont {Cagnoli}}, \bibinfo {author}
  {\bibfnamefont {J.}~\bibnamefont {Camp}}, \bibinfo {author} {\bibfnamefont
  {D.}~\bibnamefont {Crooks}}, \bibinfo {author} {\bibfnamefont {M.~M.}\
  \bibnamefont {Fejer}}, \bibinfo {author} {\bibfnamefont {A.~M.}\ \bibnamefont
  {Gretarsson}}, \bibinfo {author} {\bibfnamefont {G.~M.}\ \bibnamefont
  {Harry}}, \emph {et~al.},\ }\bibfield  {title} {\bibinfo {title} {Mechanical
  loss in tantala/silica dielectric mirror coatings},\ }\href@noop {}
  {\bibfield  {journal} {\bibinfo  {journal} {Classical and Quantum Gravity}\
  }\textbf {\bibinfo {volume} {20}},\ \bibinfo {pages} {2917} (\bibinfo {year}
  {2003})}\BibitemShut {NoStop}%
\bibitem [{\citenamefont {Granata}\ \emph {et~al.}(2016)\citenamefont
  {Granata}, \citenamefont {Saracco}, \citenamefont {Morgado}, \citenamefont
  {Cajgfinger}, \citenamefont {Cagnoli}, \citenamefont {Degallaix},
  \citenamefont {Dolique}, \citenamefont {Forest}, \citenamefont {Franc},
  \citenamefont {Michel} \emph {et~al.}}]{granata2016mechanical}%
  \BibitemOpen
  \bibfield  {author} {\bibinfo {author} {\bibfnamefont {M.}~\bibnamefont
  {Granata}}, \bibinfo {author} {\bibfnamefont {E.}~\bibnamefont {Saracco}},
  \bibinfo {author} {\bibfnamefont {N.}~\bibnamefont {Morgado}}, \bibinfo
  {author} {\bibfnamefont {A.}~\bibnamefont {Cajgfinger}}, \bibinfo {author}
  {\bibfnamefont {G.}~\bibnamefont {Cagnoli}}, \bibinfo {author} {\bibfnamefont
  {J.}~\bibnamefont {Degallaix}}, \bibinfo {author} {\bibfnamefont
  {V.}~\bibnamefont {Dolique}}, \bibinfo {author} {\bibfnamefont
  {D.}~\bibnamefont {Forest}}, \bibinfo {author} {\bibfnamefont
  {J.}~\bibnamefont {Franc}}, \bibinfo {author} {\bibfnamefont
  {C.}~\bibnamefont {Michel}}, \emph {et~al.},\ }\bibfield  {title} {\bibinfo
  {title} {Mechanical loss in state-of-the-art amorphous optical coatings},\
  }\href@noop {} {\bibfield  {journal} {\bibinfo  {journal} {Physical Review
  D}\ }\textbf {\bibinfo {volume} {93}},\ \bibinfo {pages} {012007} (\bibinfo
  {year} {2016})}\BibitemShut {NoStop}%
\bibitem [{\citenamefont {Harry}\ \emph {et~al.}(2006)\citenamefont {Harry},
  \citenamefont {Abernathy}, \citenamefont {Becerra-Toledo}, \citenamefont
  {Armandula}, \citenamefont {Black}, \citenamefont {Dooley}, \citenamefont
  {Eichenfield}, \citenamefont {Nwabugwu}, \citenamefont {Villar},
  \citenamefont {Crooks} \emph {et~al.}}]{harry2006titania}%
  \BibitemOpen
  \bibfield  {author} {\bibinfo {author} {\bibfnamefont {G.~M.}\ \bibnamefont
  {Harry}}, \bibinfo {author} {\bibfnamefont {M.~R.}\ \bibnamefont
  {Abernathy}}, \bibinfo {author} {\bibfnamefont {A.~E.}\ \bibnamefont
  {Becerra-Toledo}}, \bibinfo {author} {\bibfnamefont {H.}~\bibnamefont
  {Armandula}}, \bibinfo {author} {\bibfnamefont {E.}~\bibnamefont {Black}},
  \bibinfo {author} {\bibfnamefont {K.}~\bibnamefont {Dooley}}, \bibinfo
  {author} {\bibfnamefont {M.}~\bibnamefont {Eichenfield}}, \bibinfo {author}
  {\bibfnamefont {C.}~\bibnamefont {Nwabugwu}}, \bibinfo {author}
  {\bibfnamefont {A.}~\bibnamefont {Villar}}, \bibinfo {author} {\bibfnamefont
  {D.}~\bibnamefont {Crooks}}, \emph {et~al.},\ }\bibfield  {title} {\bibinfo
  {title} {Titania-doped tantala/silica coatings for gravitational-wave
  detection},\ }\href@noop {} {\bibfield  {journal} {\bibinfo  {journal}
  {Classical and Quantum Gravity}\ }\textbf {\bibinfo {volume} {24}},\ \bibinfo
  {pages} {405} (\bibinfo {year} {2006})}\BibitemShut {NoStop}%
\bibitem [{\citenamefont {Bassiri}\ \emph {et~al.}(2013)\citenamefont
  {Bassiri}, \citenamefont {Evans}, \citenamefont {Borisenko}, \citenamefont
  {Fejer}, \citenamefont {Hough}, \citenamefont {MacLaren}, \citenamefont
  {Martin}, \citenamefont {Route},\ and\ \citenamefont
  {Rowan}}]{bassiri2013correlations}%
  \BibitemOpen
  \bibfield  {author} {\bibinfo {author} {\bibfnamefont {R.}~\bibnamefont
  {Bassiri}}, \bibinfo {author} {\bibfnamefont {K.}~\bibnamefont {Evans}},
  \bibinfo {author} {\bibfnamefont {K.}~\bibnamefont {Borisenko}}, \bibinfo
  {author} {\bibfnamefont {M.}~\bibnamefont {Fejer}}, \bibinfo {author}
  {\bibfnamefont {J.}~\bibnamefont {Hough}}, \bibinfo {author} {\bibfnamefont
  {I.}~\bibnamefont {MacLaren}}, \bibinfo {author} {\bibfnamefont
  {I.}~\bibnamefont {Martin}}, \bibinfo {author} {\bibfnamefont
  {R.}~\bibnamefont {Route}},\ and\ \bibinfo {author} {\bibfnamefont
  {S.}~\bibnamefont {Rowan}},\ }\bibfield  {title} {\bibinfo {title}
  {{Correlations between the mechanical loss and atomic structure of amorphous
  TiO$_2$-doped Ta$_2$O$_5$ coatings}},\ }\href@noop {} {\bibfield  {journal}
  {\bibinfo  {journal} {Acta Materialia}\ }\textbf {\bibinfo {volume} {61}},\
  \bibinfo {pages} {1070} (\bibinfo {year} {2013})}\BibitemShut {NoStop}%
\bibitem [{\citenamefont {Fazio}\ \emph
  {et~al.}(2020{\natexlab{a}})\citenamefont {Fazio}, \citenamefont {Vajente},
  \citenamefont {Ananyeva}, \citenamefont {Markosyan}, \citenamefont {Bassiri},
  \citenamefont {Fejer},\ and\ \citenamefont {Menoni}}]{fazio2020structure}%
  \BibitemOpen
  \bibfield  {author} {\bibinfo {author} {\bibfnamefont {M.}~\bibnamefont
  {Fazio}}, \bibinfo {author} {\bibfnamefont {G.}~\bibnamefont {Vajente}},
  \bibinfo {author} {\bibfnamefont {A.}~\bibnamefont {Ananyeva}}, \bibinfo
  {author} {\bibfnamefont {A.}~\bibnamefont {Markosyan}}, \bibinfo {author}
  {\bibfnamefont {R.}~\bibnamefont {Bassiri}}, \bibinfo {author} {\bibfnamefont
  {M.}~\bibnamefont {Fejer}},\ and\ \bibinfo {author} {\bibfnamefont {C.~S.}\
  \bibnamefont {Menoni}},\ }\bibfield  {title} {\bibinfo {title} {{Structure
  and morphology of low mechanical loss TiO$_2$-doped Ta$_2$O$_5$}},\
  }\href@noop {} {\bibfield  {journal} {\bibinfo  {journal} {Optical Materials
  Express}\ }\textbf {\bibinfo {volume} {10}},\ \bibinfo {pages} {1687}
  (\bibinfo {year} {2020}{\natexlab{a}})}\BibitemShut {NoStop}%
\bibitem [{\citenamefont {Flaminio}\ \emph {et~al.}(2010)\citenamefont
  {Flaminio}, \citenamefont {Franc}, \citenamefont {Michel}, \citenamefont
  {Morgado}, \citenamefont {Pinard},\ and\ \citenamefont
  {Sassolas}}]{flaminio2010study}%
  \BibitemOpen
  \bibfield  {author} {\bibinfo {author} {\bibfnamefont {R.}~\bibnamefont
  {Flaminio}}, \bibinfo {author} {\bibfnamefont {J.}~\bibnamefont {Franc}},
  \bibinfo {author} {\bibfnamefont {C.}~\bibnamefont {Michel}}, \bibinfo
  {author} {\bibfnamefont {N.}~\bibnamefont {Morgado}}, \bibinfo {author}
  {\bibfnamefont {L.}~\bibnamefont {Pinard}},\ and\ \bibinfo {author}
  {\bibfnamefont {B.}~\bibnamefont {Sassolas}},\ }\bibfield  {title} {\bibinfo
  {title} {A study of coating mechanical and optical losses in view of reducing
  mirror thermal noise in gravitational wave detectors},\ }\href@noop {}
  {\bibfield  {journal} {\bibinfo  {journal} {Classical and Quantum Gravity}\
  }\textbf {\bibinfo {volume} {27}},\ \bibinfo {pages} {084030} (\bibinfo
  {year} {2010})}\BibitemShut {NoStop}%
\bibitem [{\citenamefont {Fazio}\ \emph
  {et~al.}(2020{\natexlab{b}})\citenamefont {Fazio}, \citenamefont {Yang},
  \citenamefont {Markosyan}, \citenamefont {Bassiri}, \citenamefont {Fejer},\
  and\ \citenamefont {Menoni}}]{fazio2020growth}%
  \BibitemOpen
  \bibfield  {author} {\bibinfo {author} {\bibfnamefont {M.}~\bibnamefont
  {Fazio}}, \bibinfo {author} {\bibfnamefont {L.}~\bibnamefont {Yang}},
  \bibinfo {author} {\bibfnamefont {A.}~\bibnamefont {Markosyan}}, \bibinfo
  {author} {\bibfnamefont {R.}~\bibnamefont {Bassiri}}, \bibinfo {author}
  {\bibfnamefont {M.~M.}\ \bibnamefont {Fejer}},\ and\ \bibinfo {author}
  {\bibfnamefont {C.~S.}\ \bibnamefont {Menoni}},\ }\bibfield  {title}
  {\bibinfo {title} {{Growth and characterization of Sc$_2$O$_3$ doped
  Ta$_2$O$_5$ thin films}},\ }\href@noop {} {\bibfield  {journal} {\bibinfo
  {journal} {Applied Optics}\ }\textbf {\bibinfo {volume} {59}},\ \bibinfo
  {pages} {A106} (\bibinfo {year} {2020}{\natexlab{b}})}\BibitemShut {NoStop}%
\bibitem [{\citenamefont {Yang}\ \emph {et~al.}(2020)\citenamefont {Yang},
  \citenamefont {Fazio}, \citenamefont {Vajente}, \citenamefont {Ananyeva},
  \citenamefont {Billingsley}, \citenamefont {Markosyan}, \citenamefont
  {Bassiri}, \citenamefont {Fejer},\ and\ \citenamefont
  {Menoni}}]{yang2020structural}%
  \BibitemOpen
  \bibfield  {author} {\bibinfo {author} {\bibfnamefont {L.}~\bibnamefont
  {Yang}}, \bibinfo {author} {\bibfnamefont {M.}~\bibnamefont {Fazio}},
  \bibinfo {author} {\bibfnamefont {G.}~\bibnamefont {Vajente}}, \bibinfo
  {author} {\bibfnamefont {A.}~\bibnamefont {Ananyeva}}, \bibinfo {author}
  {\bibfnamefont {G.}~\bibnamefont {Billingsley}}, \bibinfo {author}
  {\bibfnamefont {A.}~\bibnamefont {Markosyan}}, \bibinfo {author}
  {\bibfnamefont {R.}~\bibnamefont {Bassiri}}, \bibinfo {author} {\bibfnamefont
  {M.~M.}\ \bibnamefont {Fejer}},\ and\ \bibinfo {author} {\bibfnamefont
  {C.~S.}\ \bibnamefont {Menoni}},\ }\bibfield  {title} {\bibinfo {title}
  {Structural evolution that affects the room-temperature internal friction of
  binary oxide nanolaminates: Implications for ultrastable optical cavities},\
  }\href@noop {} {\bibfield  {journal} {\bibinfo  {journal} {ACS Applied Nano
  Materials}\ } (\bibinfo {year} {2020})}\BibitemShut {NoStop}%
\bibitem [{\citenamefont {Abernathy}\ \emph {et~al.}(2021)\citenamefont
  {Abernathy}, \citenamefont {Amato}, \citenamefont {Ananyeva}, \citenamefont
  {Angelova}, \citenamefont {Baloukas}, \citenamefont {Bassiri}, \citenamefont
  {Billingsley}, \citenamefont {Birney}, \citenamefont {Cagnoli}, \citenamefont
  {Canepa} \emph {et~al.}}]{abernathy2021exploration}%
  \BibitemOpen
  \bibfield  {author} {\bibinfo {author} {\bibfnamefont {M.}~\bibnamefont
  {Abernathy}}, \bibinfo {author} {\bibfnamefont {A.}~\bibnamefont {Amato}},
  \bibinfo {author} {\bibfnamefont {A.}~\bibnamefont {Ananyeva}}, \bibinfo
  {author} {\bibfnamefont {S.}~\bibnamefont {Angelova}}, \bibinfo {author}
  {\bibfnamefont {B.}~\bibnamefont {Baloukas}}, \bibinfo {author}
  {\bibfnamefont {R.}~\bibnamefont {Bassiri}}, \bibinfo {author} {\bibfnamefont
  {G.}~\bibnamefont {Billingsley}}, \bibinfo {author} {\bibfnamefont
  {R.}~\bibnamefont {Birney}}, \bibinfo {author} {\bibfnamefont
  {G.}~\bibnamefont {Cagnoli}}, \bibinfo {author} {\bibfnamefont
  {M.}~\bibnamefont {Canepa}}, \emph {et~al.},\ }\bibfield  {title} {\bibinfo
  {title} {Exploration of co-sputtered ta $_2$o$_5$-zro$_2$ thin films for
  gravitational-wave detectors},\ }\href@noop {} {\bibfield  {journal}
  {\bibinfo  {journal} {arXiv preprint arXiv:2103.14140}\ } (\bibinfo {year}
  {2021})}\BibitemShut {NoStop}%
\bibitem [{\citenamefont {Granata}\ \emph
  {et~al.}(2020{\natexlab{b}})\citenamefont {Granata}, \citenamefont {Amato},
  \citenamefont {Cagnoli}, \citenamefont {Coulon}, \citenamefont {Degallaix},
  \citenamefont {Forest}, \citenamefont {Mereni}, \citenamefont {Michel},
  \citenamefont {Pinard}, \citenamefont {Sassolas} \emph
  {et~al.}}]{granata2020progress}%
  \BibitemOpen
  \bibfield  {author} {\bibinfo {author} {\bibfnamefont {M.}~\bibnamefont
  {Granata}}, \bibinfo {author} {\bibfnamefont {A.}~\bibnamefont {Amato}},
  \bibinfo {author} {\bibfnamefont {G.}~\bibnamefont {Cagnoli}}, \bibinfo
  {author} {\bibfnamefont {M.}~\bibnamefont {Coulon}}, \bibinfo {author}
  {\bibfnamefont {J.}~\bibnamefont {Degallaix}}, \bibinfo {author}
  {\bibfnamefont {D.}~\bibnamefont {Forest}}, \bibinfo {author} {\bibfnamefont
  {L.}~\bibnamefont {Mereni}}, \bibinfo {author} {\bibfnamefont
  {C.}~\bibnamefont {Michel}}, \bibinfo {author} {\bibfnamefont
  {L.}~\bibnamefont {Pinard}}, \bibinfo {author} {\bibfnamefont
  {B.}~\bibnamefont {Sassolas}}, \emph {et~al.},\ }\bibfield  {title} {\bibinfo
  {title} {Progress in the measurement and reduction of thermal noise in
  optical coatings for gravitational-wave detectors},\ }\href@noop {}
  {\bibfield  {journal} {\bibinfo  {journal} {Applied optics}\ }\textbf
  {\bibinfo {volume} {59}},\ \bibinfo {pages} {A229} (\bibinfo {year}
  {2020}{\natexlab{b}})}\BibitemShut {NoStop}%
\bibitem [{\citenamefont {Amato}\ \emph {et~al.}(2021)\citenamefont {Amato},
  \citenamefont {Cagnoli}, \citenamefont {Granata}, \citenamefont {Sassolas},
  \citenamefont {Degallaix}, \citenamefont {Forest}, \citenamefont {Michel},
  \citenamefont {Pinard}, \citenamefont {Demos}, \citenamefont {Gras},
  \citenamefont {Evans}, \citenamefont {Di~Michele},\ and\ \citenamefont
  {Canepa}}]{PhysRevD.103.072001}%
  \BibitemOpen
  \bibfield  {author} {\bibinfo {author} {\bibfnamefont {A.}~\bibnamefont
  {Amato}}, \bibinfo {author} {\bibfnamefont {G.}~\bibnamefont {Cagnoli}},
  \bibinfo {author} {\bibfnamefont {M.}~\bibnamefont {Granata}}, \bibinfo
  {author} {\bibfnamefont {B.}~\bibnamefont {Sassolas}}, \bibinfo {author}
  {\bibfnamefont {J.}~\bibnamefont {Degallaix}}, \bibinfo {author}
  {\bibfnamefont {D.}~\bibnamefont {Forest}}, \bibinfo {author} {\bibfnamefont
  {C.}~\bibnamefont {Michel}}, \bibinfo {author} {\bibfnamefont
  {L.}~\bibnamefont {Pinard}}, \bibinfo {author} {\bibfnamefont
  {N.}~\bibnamefont {Demos}}, \bibinfo {author} {\bibfnamefont
  {S.}~\bibnamefont {Gras}}, \bibinfo {author} {\bibfnamefont {M.}~\bibnamefont
  {Evans}}, \bibinfo {author} {\bibfnamefont {A.}~\bibnamefont {Di~Michele}},\
  and\ \bibinfo {author} {\bibfnamefont {M.}~\bibnamefont {Canepa}},\
  }\bibfield  {title} {\bibinfo {title} {Optical and mechanical properties of
  ion-beam-sputtered ${\mathrm{nb}}_{2}{\mathrm{o}}_{5}$ and
  ${\mathrm{tio}}_{2}\text{\ensuremath{-}}{\mathrm{nb}}_{2}{\mathrm{o}}_{5}$
  thin films for gravitational-wave interferometers and an improved measurement
  of coating thermal noise in advanced ligo},\ }\href
  {https://doi.org/10.1103/PhysRevD.103.072001} {\bibfield  {journal} {\bibinfo
   {journal} {Phys. Rev. D}\ }\textbf {\bibinfo {volume} {103}},\ \bibinfo
  {pages} {072001} (\bibinfo {year} {2021})}\BibitemShut {NoStop}%
\bibitem [{\citenamefont {Zhurin}\ \emph {et~al.}(2000)\citenamefont {Zhurin},
  \citenamefont {Kaufman}, \citenamefont {Kahn},\ and\ \citenamefont
  {Hylton}}]{zhurin2000biased}%
  \BibitemOpen
  \bibfield  {author} {\bibinfo {author} {\bibfnamefont {V.}~\bibnamefont
  {Zhurin}}, \bibinfo {author} {\bibfnamefont {H.}~\bibnamefont {Kaufman}},
  \bibinfo {author} {\bibfnamefont {J.}~\bibnamefont {Kahn}},\ and\ \bibinfo
  {author} {\bibfnamefont {T.}~\bibnamefont {Hylton}},\ }\bibfield  {title}
  {\bibinfo {title} {Biased target deposition},\ }\href@noop {} {\bibfield
  {journal} {\bibinfo  {journal} {Journal of Vacuum Science \& Technology A:
  Vacuum, Surfaces, and Films}\ }\textbf {\bibinfo {volume} {18}},\ \bibinfo
  {pages} {37} (\bibinfo {year} {2000})}\BibitemShut {NoStop}%
\bibitem [{\citenamefont {Alexandrovski}\ \emph {et~al.}(2009)\citenamefont
  {Alexandrovski}, \citenamefont {Fejer}, \citenamefont {Markosian},\ and\
  \citenamefont {Route}}]{alexandrovski2009photothermal}%
  \BibitemOpen
  \bibfield  {author} {\bibinfo {author} {\bibfnamefont {A.}~\bibnamefont
  {Alexandrovski}}, \bibinfo {author} {\bibfnamefont {M.}~\bibnamefont
  {Fejer}}, \bibinfo {author} {\bibfnamefont {A.}~\bibnamefont {Markosian}},\
  and\ \bibinfo {author} {\bibfnamefont {R.}~\bibnamefont {Route}},\ }\bibfield
   {title} {\bibinfo {title} {Photothermal common-path interferometry (pci):
  new developments},\ }in\ \href@noop {} {\emph {\bibinfo {booktitle} {Solid
  State Lasers XVIII: Technology and Devices}}},\ Vol.\ \bibinfo {volume}
  {7193}\ (\bibinfo {organization} {International Society for Optics and
  Photonics},\ \bibinfo {year} {2009})\ p.\ \bibinfo {pages}
  {71930D}\BibitemShut {NoStop}%
\bibitem [{\citenamefont {Fairley}(2005)}]{casaxps}%
  \BibitemOpen
  \bibfield  {author} {\bibinfo {author} {\bibfnamefont {M.}~\bibnamefont
  {Fairley}},\ }\href@noop {} {\emph {\bibinfo {title} {$\textcopyright$ Casa
  Software Ltd.}}}\ (\bibinfo {year} {2005})\BibitemShut {NoStop}%
\bibitem [{\citenamefont {Cesarini}\ \emph {et~al.}(2009)\citenamefont
  {Cesarini}, \citenamefont {Lorenzini}, \citenamefont {Campagna},
  \citenamefont {Martelli}, \citenamefont {Piergiovanni}, \citenamefont
  {Vetrano}, \citenamefont {Losurdo},\ and\ \citenamefont
  {Cagnoli}}]{cesarini2009}%
  \BibitemOpen
  \bibfield  {author} {\bibinfo {author} {\bibfnamefont {E.}~\bibnamefont
  {Cesarini}}, \bibinfo {author} {\bibfnamefont {M.}~\bibnamefont {Lorenzini}},
  \bibinfo {author} {\bibfnamefont {E.}~\bibnamefont {Campagna}}, \bibinfo
  {author} {\bibfnamefont {F.}~\bibnamefont {Martelli}}, \bibinfo {author}
  {\bibfnamefont {F.}~\bibnamefont {Piergiovanni}}, \bibinfo {author}
  {\bibfnamefont {F.}~\bibnamefont {Vetrano}}, \bibinfo {author} {\bibfnamefont
  {G.}~\bibnamefont {Losurdo}},\ and\ \bibinfo {author} {\bibfnamefont
  {G.}~\bibnamefont {Cagnoli}},\ }\bibfield  {title} {\bibinfo {title} {A
  “gentle” nodal suspension for measurements of the acoustic attenuation in
  materials},\ }\href {https://doi.org/10.1063/1.3124800} {\bibfield  {journal}
  {\bibinfo  {journal} {Review of Scientific Instruments}\ }\textbf {\bibinfo
  {volume} {80}},\ \bibinfo {pages} {053904} (\bibinfo {year}
  {2009})}\BibitemShut {NoStop}%
\bibitem [{\citenamefont {Vajente}\ \emph {et~al.}(2017)\citenamefont
  {Vajente}, \citenamefont {Ananyeva}, \citenamefont {Billingsley},
  \citenamefont {Gustafson}, \citenamefont {Heptonstall}, \citenamefont
  {Sanchez},\ and\ \citenamefont {Torrie}}]{vajente2017}%
  \BibitemOpen
  \bibfield  {author} {\bibinfo {author} {\bibfnamefont {G.}~\bibnamefont
  {Vajente}}, \bibinfo {author} {\bibfnamefont {A.}~\bibnamefont {Ananyeva}},
  \bibinfo {author} {\bibfnamefont {G.}~\bibnamefont {Billingsley}}, \bibinfo
  {author} {\bibfnamefont {E.}~\bibnamefont {Gustafson}}, \bibinfo {author}
  {\bibfnamefont {A.}~\bibnamefont {Heptonstall}}, \bibinfo {author}
  {\bibfnamefont {E.}~\bibnamefont {Sanchez}},\ and\ \bibinfo {author}
  {\bibfnamefont {C.}~\bibnamefont {Torrie}},\ }\bibfield  {title} {\bibinfo
  {title} {A high throughput instrument to measure mechanical losses in thin
  film coatings},\ }\href {https://doi.org/10.1063/1.4990036} {\bibfield
  {journal} {\bibinfo  {journal} {Review of Scientific Instruments}\ }\textbf
  {\bibinfo {volume} {88}},\ \bibinfo {pages} {073901} (\bibinfo {year}
  {2017})}\BibitemShut {NoStop}%
\bibitem [{\citenamefont {Rao}(2007)}]{rao2007}%
  \BibitemOpen
  \bibfield  {author} {\bibinfo {author} {\bibfnamefont {S.~S.}\ \bibnamefont
  {Rao}},\ }\href {https://doi.org/10.1002/9780470117866} {\emph {\bibinfo
  {title} {Vibration of Continuous Systems}}}\ (\bibinfo  {publisher} {John
  Wiley and Sons},\ \bibinfo {year} {2007})\BibitemShut {NoStop}%
\bibitem [{\citenamefont {Vajente}\ \emph {et~al.}(2020)\citenamefont
  {Vajente}, \citenamefont {Fazio}, \citenamefont {Yang}, \citenamefont
  {Gupta}, \citenamefont {Ananyeva}, \citenamefont {Billinsley},\ and\
  \citenamefont {Menoni}}]{vajente2020}%
  \BibitemOpen
  \bibfield  {author} {\bibinfo {author} {\bibfnamefont {G.}~\bibnamefont
  {Vajente}}, \bibinfo {author} {\bibfnamefont {M.}~\bibnamefont {Fazio}},
  \bibinfo {author} {\bibfnamefont {L.}~\bibnamefont {Yang}}, \bibinfo {author}
  {\bibfnamefont {A.}~\bibnamefont {Gupta}}, \bibinfo {author} {\bibfnamefont
  {A.}~\bibnamefont {Ananyeva}}, \bibinfo {author} {\bibfnamefont
  {G.}~\bibnamefont {Billinsley}},\ and\ \bibinfo {author} {\bibfnamefont
  {C.~S.}\ \bibnamefont {Menoni}},\ }\bibfield  {title} {\bibinfo {title}
  {Method for the experimental measurement of bulk and shear loss angles in
  amorphous thin films},\ }\href {https://doi.org/10.1103/PhysRevD.101.042004}
  {\bibfield  {journal} {\bibinfo  {journal} {Phys. Rev. D}\ }\textbf {\bibinfo
  {volume} {101}},\ \bibinfo {pages} {042004} (\bibinfo {year}
  {2020})}\BibitemShut {NoStop}%
\bibitem [{\citenamefont {Hong}\ \emph {et~al.}(2013)\citenamefont {Hong},
  \citenamefont {Yang}, \citenamefont {Gustafson}, \citenamefont {Adhikari},\
  and\ \citenamefont {Chen}}]{Hong2013}%
  \BibitemOpen
  \bibfield  {author} {\bibinfo {author} {\bibfnamefont {T.}~\bibnamefont
  {Hong}}, \bibinfo {author} {\bibfnamefont {H.}~\bibnamefont {Yang}}, \bibinfo
  {author} {\bibfnamefont {E.~K.}\ \bibnamefont {Gustafson}}, \bibinfo {author}
  {\bibfnamefont {R.~X.}\ \bibnamefont {Adhikari}},\ and\ \bibinfo {author}
  {\bibfnamefont {Y.}~\bibnamefont {Chen}},\ }\bibfield  {title} {\bibinfo
  {title} {Brownian thermal noise in multilayer coated mirrors},\ }\href
  {https://doi.org/10.1103/PhysRevD.87.082001} {\bibfield  {journal} {\bibinfo
  {journal} {Phys. Rev. D}\ }\textbf {\bibinfo {volume} {87}},\ \bibinfo
  {pages} {082001} (\bibinfo {year} {2013})}\BibitemShut {NoStop}%
\bibitem [{\citenamefont {Zallen}(2008)}]{zallen2008physics}%
  \BibitemOpen
  \bibfield  {author} {\bibinfo {author} {\bibfnamefont {R.}~\bibnamefont
  {Zallen}},\ }\href@noop {} {\emph {\bibinfo {title} {The physics of amorphous
  solids}}}\ (\bibinfo  {publisher} {John Wiley \& Sons},\ \bibinfo {year}
  {2008})\BibitemShut {NoStop}%
\bibitem [{\citenamefont {Yang}\ \emph {et~al.}(2021)\citenamefont {Yang},
  \citenamefont {Vajente}, \citenamefont {Fazio}, \citenamefont {Ananyeva},
  \citenamefont {Billingsley}, \citenamefont {Markosyan}, \citenamefont
  {Bassiri}, \citenamefont {Prasai}, \citenamefont {Fejer},\ and\ \citenamefont
  {Menoni}}]{yang2021enhanced}%
  \BibitemOpen
  \bibfield  {author} {\bibinfo {author} {\bibfnamefont {L.}~\bibnamefont
  {Yang}}, \bibinfo {author} {\bibfnamefont {G.}~\bibnamefont {Vajente}},
  \bibinfo {author} {\bibfnamefont {M.}~\bibnamefont {Fazio}}, \bibinfo
  {author} {\bibfnamefont {A.}~\bibnamefont {Ananyeva}}, \bibinfo {author}
  {\bibfnamefont {G.}~\bibnamefont {Billingsley}}, \bibinfo {author}
  {\bibfnamefont {A.}~\bibnamefont {Markosyan}}, \bibinfo {author}
  {\bibfnamefont {R.}~\bibnamefont {Bassiri}}, \bibinfo {author} {\bibfnamefont
  {K.}~\bibnamefont {Prasai}}, \bibinfo {author} {\bibfnamefont {M.~M.}\
  \bibnamefont {Fejer}},\ and\ \bibinfo {author} {\bibfnamefont {C.~S.}\
  \bibnamefont {Menoni}},\ }\bibfield  {title} {\bibinfo {title} {Enhanced
  medium range order in vapor deposited germania glasses at elevated
  temperatures},\ }\href@noop {} {\bibfield  {journal} {\bibinfo  {journal}
  {arXiv preprint arXiv:2102.08526}\ } (\bibinfo {year} {2021})}\BibitemShut
  {NoStop}%
\bibitem [{\citenamefont {Fazio}\ \emph {et~al.}(2021)\citenamefont {Fazio},
  \citenamefont {Yang},\ and\ \citenamefont {Menoni}}]{fazio2021prediction}%
  \BibitemOpen
  \bibfield  {author} {\bibinfo {author} {\bibfnamefont {M.~A.}\ \bibnamefont
  {Fazio}}, \bibinfo {author} {\bibfnamefont {L.}~\bibnamefont {Yang}},\ and\
  \bibinfo {author} {\bibfnamefont {C.~S.}\ \bibnamefont {Menoni}},\ }\bibfield
   {title} {\bibinfo {title} {{Prediction of crystallized phases of amorphous
  Ta$_2$O$_5$-based mixed oxide thin films using a density functional theory
  database}},\ }\href@noop {} {\bibfield  {journal} {\bibinfo  {journal} {APL
  Materials}\ }\textbf {\bibinfo {volume} {9}},\ \bibinfo {pages} {031106}
  (\bibinfo {year} {2021})}\BibitemShut {NoStop}%
\bibitem [{\citenamefont {Gates-Rector}\ and\ \citenamefont
  {Blanton}(2019)}]{gates2019powder}%
  \BibitemOpen
  \bibfield  {author} {\bibinfo {author} {\bibfnamefont {S.}~\bibnamefont
  {Gates-Rector}}\ and\ \bibinfo {author} {\bibfnamefont {T.}~\bibnamefont
  {Blanton}},\ }\bibfield  {title} {\bibinfo {title} {The powder diffraction
  file: a quality materials characterization database},\ }\href@noop {}
  {\bibfield  {journal} {\bibinfo  {journal} {Powder Diffraction}\ }\textbf
  {\bibinfo {volume} {34}},\ \bibinfo {pages} {352} (\bibinfo {year}
  {2019})}\BibitemShut {NoStop}%
\end{thebibliography}%

\end{document}